\documentclass[11pt]{article}
\usepackage
[
        a4paper,
        left=1in,
        right=1in,
        top=1in,
        bottom=1in
]
{geometry}
\usepackage{amsmath,amssymb,amsthm}
\usepackage{hyperref}
\usepackage{graphicx}

\usepackage{xcolor}

\definecolor{defblue}{rgb}{0.121,0.47,0.705}
\DeclareTextFontCommand{\emph}{\color{defblue}\em}

\usepackage[most,many,breakable]{tcolorbox}

\usepackage{varwidth}
\definecolor{lipicsblue}{rgb}{0.08235294118,0.3098039216,0.537254902}
\definecolor{lipicsyellow}{HTML}{e6b919}
\definecolor{linkblue}{rgb}{0.098,0.098,0.4392}
\definecolor{ourgreen}{rgb}{0.509,0.745,0.235}
\definecolor{indianred}{rgb}{0.804,0.361,0.361}
\definecolor{indianred1}{rgb}{1,0.416,0.416}
\definecolor{indianred3}{rgb}{0.804,0.333,0.333}
\definecolor{orangered}{rgb}{1,0.271,0}
\definecolor{coral1}{rgb}{1,0.447,0.337}
\definecolor{rosybrown2}{rgb}{0.933,0.231,0.231}
\definecolor{aquamarine4}{rgb}{0.271,0.545,0.455}
\definecolor{chartreuse3}{rgb}{0.4,0.804,0}
\definecolor{mediumpurple3}{rgb}{0.537,0.408,0.804}
\definecolor{mediumvioletred}{rgb}{0.78,0.082, 0.522}
\hypersetup{colorlinks=true,
    linkcolor=orangered,
    anchorcolor=lipicsblue,
    citecolor=lipicsblue,
    filecolor=lipicsblue,
    menucolor=lipicsblue,
    urlcolor=lipicsblue,
    bookmarksopen=true,
    bookmarksopenlevel=2,
    bookmarksnumbered=true,
    plainpages=false,
    }

\newtcolorbox{Definition}[2]{enhanced,
	before skip=2mm,after skip=2mm, colback=lipicsyellow!5!white,colframe=lipicsyellow!100!white,boxrule=0.5mm,
	attach boxed title to top left={xshift=1cm,yshift*=1mm-\tcboxedtitleheight}, varwidth boxed title*=-3cm,
	boxed title style={frame code={
					\path[fill=tcbcolback]
					([yshift=-1mm,xshift=-1mm]frame.north west)
					arc[start angle=0,end angle=180,radius=1mm]
					([yshift=-1mm,xshift=1mm]frame.north east)
					arc[start angle=180,end angle=0,radius=1mm];
					\path[left color=tcbcolback!60!white,right color=tcbcolback!60!white,
						middle color=tcbcolback!80!white]
					([xshift=-2mm]frame.north west) -- ([xshift=2mm]frame.north east)
					[rounded corners=1mm]-- ([xshift=1mm,yshift=-1mm]frame.north east)
					-- (frame.south east) -- (frame.south west)
					-- ([xshift=-1mm,yshift=-1mm]frame.north west)
					[sharp corners]-- cycle;
				},interior engine=empty,
		},
	fonttitle=\bfseries,
colbacktitle=lipicsyellow!100!white, title={#1}}

\newcommand{\dfn}[2]{\begin{Definition}{#1}{}#2\end{Definition}}

\definecolor{myg}{RGB}{56, 140, 70}
\tcbuselibrary{theorems,skins,hooks}
\newtcbtheorem[number within=section]{Theorem}{DP}
{%
	enhanced,
	breakable,
	colback = myg!10,
	frame hidden,
	boxrule = 0sp,
	borderline west = {2pt}{0pt}{myg},
	sharp corners,
	detach title,
	before upper = \tcbtitle\par\smallskip,
	coltitle = myg!85!black,
	fonttitle = \bfseries\sffamily,
	description font = \mdseries,
	separator sign none,
	segmentation style={solid, myg},
    title={#1}
}
{th}

\newcommand{\thm}[2]{\begin{Theorem}{#1}{}#2\end{Theorem}}

\newtheorem{remark}{Remark}

\tcbuselibrary{theorems,skins,hooks}
\newtcbtheorem[number within=section]{Complexity}{Complexity}
{%
	enhanced,
	breakable,
	colback = lipicsyellow!10,
	frame hidden,
	boxrule = 0sp,
	borderline west = {2pt}{0pt}{lipicsyellow},
	sharp corners,
	detach title,
	before upper = \tcbtitle\par\smallskip,
	coltitle = lipicsyellow!85!black,
	fonttitle = \bfseries\sffamily,
	description font = \mdseries,
	separator sign none,
	segmentation style={solid, lipicsyellow},
    title={#1}
}
{th}

\newcommand{\runningtime}[2]{\begin{Complexity}{#1}{}#2\end{Complexity}}

\usepackage{graphicx} %
\usepackage{hyperref}
\usepackage{todonotes}
\usepackage{amsmath}
\usepackage{caption}
\usepackage{xspace}
\usepackage{todonotes}
\usepackage{complexity}
\usepackage{braket}
\usepackage{amsmath,amsfonts,amssymb}
\newcommand{\myceil}[1]{\ensuremath{\lceil #1 \rceil}}

\usepackage{algorithm}
\usepackage{algpseudocode}
\usepackage[linewidth=1pt]{mdframed}
\usepackage{breqn}
\usepackage[inline]{enumitem}
\usepackage{subcaption}

\DeclareMathOperator{\operator}{op}
\DeclareMathOperator{\Time}{Time}
\DeclareMathOperator{\operatorBF}{\mathbf{op}}

\DeclareMathOperator{\find}{find}
\DeclareMathOperator{\findAll}{findAll}

\DeclareMathOperator{\solution}{sol}
\newcommand{\depgr}{\ensuremath{G_{\mathcal{P}}(I)}}

\newtheorem{theorem}{Theorem}
\usepackage{url} 
\usepackage{hyperref}
\usepackage[capitalise,nameinlink]{cleveref}

\Crefname{observation}{Observation}{Observations}
\Crefname{algorithm}{Algorithm}{Algorithms}
\Crefname{algocf}{Algorithm}{Algorithms}
\Crefname{section}{Section}{Sections}
\Crefname{lemma}{Lemma}{Lemmata}
\Crefname{note}{Note}{Notes}
\Crefname{claim}{Claim}{Claims}
\Crefname{property}{Property}{Properties}
\Crefname{enumi}{Property}{Properties}
\Crefname{figure}{Figure}{Figures}
\Crefname{fact}{Fact}{Facts}
\Crefname{equation}{Equation}{Equations}
 \usepackage{multirow}
\newcommand{\bigO}{O}
\newcommand{\tildeO}{\tilde{\bigO}}

\usepackage{xcolor,soul}

\usepackage{framed}
\usepackage{tabularx}

\usepackage{tikz}
\usetikzlibrary{calc}

\newlength{\RoundedBoxWidth}
\newsavebox{\GrayRoundedBox}
\newenvironment{GrayBox}[1]%
   {\setlength{\RoundedBoxWidth}{0.95\columnwidth}
    \def\boxheading{#1}
    \begin{lrbox}{\GrayRoundedBox}
       \begin{minipage}{\RoundedBoxWidth}}%
   {   \end{minipage}
    \end{lrbox}
    \begin{center}
    \begin{tikzpicture}%
       \node(Text)[draw=black!20,fill=white,rounded corners,inner xsep=2ex,inner ysep=2ex,text width=\RoundedBoxWidth]
             {\usebox{\GrayRoundedBox}};
        \coordinate(x) at (current bounding box.north west);
        \node [draw=white,rectangle,inner sep=3pt,anchor=north west,fill=white]
        at ($(x)+(6pt,.75em)$) {\boxheading};
    \end{tikzpicture}
    \end{center}}

\newenvironment{defproblemx}[2]{\noindent\ignorespaces%
                                \FrameSep=6pt%
                                \parindent=6pt
                \vspace{-3mm}            
                \begin{GrayBox}{#1}%
                \begin{tabular*}{0.98\columnwidth}{!{\extracolsep{\fill}}@{} >{\itshape} p{#2} p{0.87\columnwidth} @{\hspace{.5em}}}%
            }{\\[-1.5ex]
                \end{tabular*}%
                \end{GrayBox}%
                \ignorespacesafterend
                \vspace{-4mm}
            }

\newcommand{\problemQuestionOutput}[3]{%
  \begin{defproblemx}{#1}{0.5cm}
    Input: & #2 \\
    Output: & #3
  \end{defproblemx}
}

\usepackage{authblk}

\date{}

\title{Quantum Speedups for Polynomial-Time\\Dynamic Programming Algorithms}

\author[1]{\href{mailto:susanna.caroppo@uniroma3.it}{Susanna Caroppo}}
\author[1]{\href{mailto:giordano.dalozzo@uniroma3.it}{Giordano Da Lozzo}}
\author[1]{\href{mailto:giuseppe.dibattista@uniroma3.it}{Giuseppe Di Battista}}
\author[2]{\href{mailto:goodrich@uci.edu}{Michael T. Goodrich}}
\author[3]{\href{mailto:noellenburg@ac.tuwien.ac.at}{Martin Nöllenburg}}

\affil[1]{Roma Tre University, Rome, Italy}
\affil[2]{University of California, Irvine, USA}
\affil[3]{TU Wien, Vienna, Austria}

\begin{document}

\maketitle

\begin{abstract}
\noindent We introduce a quantum dynamic programming framework that allows us to directly extend to the quantum realm a large body of classical dynamic programming algorithms. The corresponding {\bf \em quantum dynamic programming algorithms} retain the same space complexity as their classical counterpart, while achieving a computational speedup. 
For a combinatorial ({\em search} or {\em optimization}) problem $\mathcal P$ and an instance $I$ of $\mathcal P$, such a speedup can be expressed in terms of the average degree $\delta$  of the {dependency digraph} $\depgr$ of $I$, determined by a recursive formulation of $\mathcal P$. The nodes of this graph are the subproblems of $\mathcal P$ induced by $I$ and its arcs are directed from each subproblem to those on whose solution it relies. In particular, our framework allows us to solve the considered problems in $\tildeO(|V(\depgr)| \sqrt{\delta})$ time. As an example, we obtain a quantum version of the Bellman-Ford algorithm for computing shortest paths from a single source vertex to all the other vertices in a weighted $n$-vertex digraph with $m$ edges that runs in $\tildeO(n\sqrt{nm})$ time, which improves the best known classical upper bound when $m \in \Omega(n^{1.4})$.
\end{abstract}

\section{Introduction}

Quantum computing represents a paradigm shift in computation, leveraging the unique principles of quantum mechanics---superposition, entanglement, and interference---to solve problems that are intractable for classical computers. These principles allow quantum algorithms to achieve significant speedups for tasks such as factoring large numbers~\cite{DBLP:journals/siamrev/Shor99}, searching unsorted databases~\cite{DBLP:conf/stoc/Grover96}, and simulating complex physical systems~\cite{RevModPhys.86.153}.  Classical computing has introduced fundamental algorithmic design paradigms that enable the efficient solution of combinatorial problems. Among these, for problems that exhibit a recursive structure, \emph{dynamic programming} and \emph{divide and conquer}, stand out predominantly for their efficiency and wide applicability. In this research, we introduce a framework that extends classical dynamic programming algorithms~\cite{DBLP:journals/cacm/Bellman61a,DBLP:journals/jgaa/BenkertHKN09,DBLP:books/daglib/0023376,DBLP:books/daglib/0017733,
DBLP:books/x/E2019,DBLP:conf/isaac/GemsaNN13,DBLP:journals/jda/HanT16,DBLP:books/daglib/0016921,DBLP:conf/soda/HuangJQ25,DBLP:books/daglib/0004298,DBLP:books/daglib/0015106,KLINCSEK1980121} to faster quantum counterparts.  This framework applies to combinatorial search and optimization problems tackled by computing dynamic programming tables.

\noindent{\bf Dynamic Programming.}
Let $\mathcal{P}$ be a combinatorial problem and let $I$ be an instance of $\mathcal{P}$ of size $n$. 
The {\bf dynamic programming paradigm} for algorithm design can often be applied to compute a solution $\solution(I)$ of $I$ for $\mathcal{P}$, if $\mathcal{P}$ satisfies the \emph{optimal substructure property}, i.e., an optimal solution for $I$ can be decomposed into (interchangeable) optimal solutions for subinstances of $I$, and the recursive computations of solutions for larger instances require solutions for \emph{overlapping subproblems}---unlike most divide-and-conquer algorithms, in which the subproblems do not overlap.
The underlying idea of (bottom-up) dynamic programming is to create a table $D$, which stores the (values of) optimal solutions for all relevant subinstances of $I$, starting from the smallest ones and then computing optimal solutions for larger and larger subinstances by suitably combining the existing solutions of smaller subinstances.
Often, the construction of $\solution(I)$, given $D$, is a simple task\footnote{The solution $\solution(I)$ of the actual instance $I$ is frequently found in the ``last'' cell of $D$. Such entry is later denoted as $D[i^*_1][i^*_2]\dots[i^*_k]$.}. Therefore, the time and space complexity for solving $\mathcal P$ are asymptotically bounded by those of constructing and storing $D$, respectively. In particular, the construction time of $D$ can be easily upper bounded by multiplying its size by the time required to recursively compute each table entry. For instance, a table $D$ of size $n^2$ and a linear-time computation for each entry yield a running time of $O(n^3)$.

\noindent{\bf Quantum Dynamic Programming.}
Quantum dynamic programming algorithms have recently attracted considerable interest in the exponential-time regime (which we review later). Surprisingly, instead, polynomial-time quantum algorithms have not yet received as much attention, despite the potential for significant speedups in practical applications. 
Khadiev and Safina \cite{DBLP:conf/uc/KhadievS19} proposed polynomial-time quantum dynamic programming algorithms for solving problems on directed acyclic graphs (DAGs).
Furrow~\cite{DBLP:journals/qic/Furrow08} gives a polynomial-time quantum dynamic programming algorithm for the {\sc Coin Change} problem and for the {\sc Maximum Subarray Sum}\footnote{It is worth remarking that, as discussed in~\cite{DBLP:journals/qic/Furrow08}, the proposed algorithm, albeit based on computing an auxiliary table, is not solved via dynamic programming, but rather by following a greedy approach.} problem.

The pioneering work by Ambainis et al.~\cite{DBLP:conf/soda/AmbainisBIKPV19} introduced two quantum dynamic programming frameworks that provide a novel approach for speeding up some 
{\bf exponential-time} classical dynamic programming algorithms, developed to address NP-complete problems. Both frameworks use classical computation to construct partial dynamic programming tables, stored in QRAM, to be accessed via quantum search subroutines. 
The first, and simpler, framework addresses a subclass of set problems where the solution for a set $S$ of size $n$ can be determined by considering all partitions of $S$ into two sets of sizes $k$ and $n-k$ (for any positive integer $k$), and by selecting the optimal option. In this setting, a quantum advantage is obtained by suitably selecting the subset of the entries to be classically precomputed so as to balance the time needed to compute the solution (obtained by identifying optimal combinations of subsets) to the problem by performing (bounded-depth) recursive quantum search primitives. For instance, this framework allows to obtain fast exponential-time quantum algorithms for {\sc Traveling Salesman}~\cite{DBLP:conf/soda/AmbainisBIKPV19}, for {\sc One-Sided Crossing Minimization}~\cite{DBLP:conf/gd/CaroppoLB24,CAROPPO2025115424}\footnote{A quantum algorithm for {\sc Two-Sided Crossing Minimization}, solely based on Grover's search, has been presented in \cite{DBLP:conf/walcom/CaroppoLB24,DBLP:journals/jgaa/CaroppoLB25}.}, for {\sc Graph Coloring}~\cite{DBLP:journals/algorithmica/ShimizuM22}, and for {\sc Dynamic programming across the subset}~\cite{DBLP:journals/anor/TkindtCL24}.
The second, more complex, framework is based on the ability of efficiently solving a \emph{Path in the Hypercube} (PHC) problem. Given a subgraph $G$ of the Boolean hypercube, where edges are directed from vertices of lower Hamming weight to vertices of higher Hamming weight, the problem asks to determine if there exists a directed path in $G$ from the vertex $0^n$ to the vertex $1^n$. In this setting, a quantum advantage is reached by combining the access to precomputed partial dynamic programming tables, with recursive applications of the quantum algorithm solving PHC. For instance, this framework allows to obtain fast exponential-time quantum algorithms for {\sc Domatic Number}~\cite{DBLP:journals/bjmc/AmbainisR24} and for {\sc Treewidth Computation}~\cite{DBLP:conf/tqc/KlevickisPV22}. Furthermore, Glos et al.~\cite{DBLP:conf/mfcs/GlosKMV21} generalized this framework to a quantum algorithm for finding a path in $n$-dimensional lattice graphs.

\noindent{\bf Our contributions.}
While exponential-time quantum algorithms aim to tackle problems beyond the reach of classical computation, polynomial-time quantum algorithms can provide substantial efficiency gains for problems already considered tractable.
In this work, we focus on polynomial-time quantum algorithms and introduce a framework that systematically extends classical dynamic programming algorithms into accelerated quantum counterparts, by harnessing quantum parallelism and amplitude amplification.
Specifically, we developed quantum subroutines that integrate quantum search  primitives, such as {\em min}, {\em max}, {\em find}, and {\em findAll}, within a dynamic programming context. These subroutines construct a superposition enabling access to the specific subset of previously computed entries, stored in a Quantum Random Access Memory (QRAM), that is needed for computing the currently-considered entry. This, in turn, enables us to harness the full potential of quantum search primitives by restricting the search space to the relevant candidates.

To demonstrate the broad usability of our framework, we apply it to several well-known problems for which the best worst-case classical algorithms rely on dynamic programming; see \cref{tab:complexities}. In many cases, applying
our framework requires additional interesting 
data structures or algorithmic steps
that are not used in corresponding classical algorithms, but, 
for space reasons, we discuss in the main text only the application to the {\sc Single-Source Shortest Path} problem and discuss the application to the remaining problems in \cref{se:appendix-uniform}.

\begin{table}[tb!]
    \centering
    \renewcommand{\arraystretch}{1.4}
    \resizebox{\columnwidth}{!}{%
    \begin{tabular}{|l|c|c|c|}
        \hline
        \textbf{Problem} & \textbf{$\operatorBF$} & \textbf{Classical Complexity} & \textbf{Quantum Complexity} \\
        \hline
        Minimum-Weight Triangulation of Convex Polygon~\cref{ssc:MWT} & $\min$ & $\bigO(n^3)$ \cite{KLINCSEK1980121} & $\tildeO(n^2\sqrt{n})$ \\
        All-Pairs Shortest Paths~\cref{ssc: APSP_Matrix} & $\min$ & $\bigO(n^3 \log \log n / \log ^2 n)$~\cite{DBLP:journals/jda/HanT16} & $\tildeO(n^2\sqrt{n}\log n)$ \\
        Single-Source Shortest Paths~\cref{sc:Bellman-Ford}& $\min$ & $\tildeO(mn^{4/5})$ \cite{DBLP:conf/soda/HuangJQ25} & $\tildeO(n\sqrt{nm})$  \\
        \multirow{2}{*}{Multi-Criteria Boundary Labeling~\cref{ssc: MCBL}}
        & $\min$ & $po$-leaders  $\bigO(n^3)$ \cite{DBLP:journals/jgaa/BenkertHKN09} & $\tildeO(n^2\sqrt{n})$  \\
        & $\min$ & $do$-leaders  $\bigO(n^5)$ \cite{DBLP:journals/jgaa/BenkertHKN09} & $\tildeO(n^4\sqrt{n})$  \\
        Segmented Least Squares~\cref{ssc: SLS} & $\min$ & $\bigO(n^2)$ \cite{DBLP:journals/cacm/Bellman61a, DBLP:books/daglib/0015106} & $\tildeO(n\sqrt{n})$ \\
        \hline
        RNA Secondary Structure~\cref{ssc: RNA-SS} & $\max$ & $\bigO(n^3)$ \cite{DBLP:books/daglib/0015106} & $\tildeO(n^2\sqrt{n})$ \\
        Rod Cutting~\cref{ssc: RC} & $\max$ & $\bigO(n^2)$ \cite{DBLP:books/daglib/0023376} & $\tildeO(n\sqrt{n})$ \\
        Largest Divisible Subset in Array~\cref{ssc: LDS} & $\max$ & $\bigO(n^2)~\cite{LDS}$ & $\tildeO(n\sqrt{n})$ \\
        Unbounded Knapsack~\cref{ssc: UKP} & $\max$ & $\bigO(Wn)$~\cite{DBLP:books/daglib/0017733} & $\tildeO(W\sqrt{n})$ \\
        Viterbi Path Problem~\cref{ssc: VA} & $\max$ & $\bigO(T \times |S|^2)$ \cite{DBLP:books/daglib/0004298} & $\tildeO(T \times |S| \times \sqrt{|S|})$ \\
        \hline
        Text Segmentation~\cref{ssc: TSP} & $\text{find}$ & $\bigO(n^2)$~\cite{DBLP:books/x/E2019} & $\tildeO(n\sqrt{n})$ \\
        Membership in Context-Free Language (CYK) \cref{sc:CYK} & $\text{findAll}$ & $\bigO(n^{2.37})$ \cite{DBLP:conf/soda/WilliamsXXZ24} & $\tildeO(n^2\sqrt{n})$ \\
        \hline
    \end{tabular}
    }
    \caption{Comparison of classical and quantum time complexities, classified based on the operator $\operator \in \{\min,\max,\find,\findAll\}$ in their recursive fomulation.}
    \label{tab:complexities}
\end{table}

\section{Preliminaries}\label{se:preliminaries}

In this section, we introduce preliminary notation and definitions.

\noindent{\bf Notation.}
Given two $k$-tuples of integers, $\langle a_1, a_2, \dots, a_k\rangle$ and $\langle b_1, b_2, \dots, b_k\rangle$, we say that $\langle a_1, a_2, \dots, a_k\rangle$ \emph{lexicographically precedes} $\langle b_1, b_2, \dots, b_k\rangle$, denoted by  $\langle a_1, a_2, \dots, a_k\rangle \prec \linebreak \langle b_1, b_2, \dots, b_k\rangle$, if there exists an index $j \in \{1, 2, \dots, k\}$ such that $a_i = b_i$ for all $i < j$, and $a_j < b_j$. The relation $\prec$ defines a total order among the $k$-tuples of integers.
Given a directed graph $G=(V,E)$, we denote by $\deg(v)$ the \emph{degree} of a vertex $v$ of $G$, that is, the number of arcs in $E$ having $v$ as their tail or head. Moreover, we denote by $\deg_{out}(v)$ the \emph{outdegree} of a vertex $v$ of $G$, that is, the number of arcs in $E$ having $v$ as their tail. The average degree of $G$ is $\delta=\frac{1}{n}\sum_{v\in V(G)}\deg(v) = \frac{2m}{n}$.
In order to simplify the notation, we use $[h]$ to denote the set $\{0,\dots,h-1\}$, where $h$ is a positive integer.
Also, given positive integers $a$ and $b$, we denote $\myceil{\frac{a}{b}}$ as $\frac{a}{b}$ and $\myceil{\log a}$ as $\log a$. 
If $f(n)=\bigO(\log^c n)$ for some constant $c$, we write $f(n)=polylog(n)$. In case $f(n)={\bigO}(n^d polylog(n))$ for some constant $d$, we use the notation $f(n)=\tildeO(n^d)$ (see, e.g.,~\cite{DBLP:journals/dam/Woeginger08}).

\noindent{\bf Combinatorial problems.}
A \emph{combinatorial search problem} $\mathcal{P}$ is a triple $\langle\Lambda, S, R\rangle$, where:
\begin{itemize}
    \item $\Lambda$ is the set of \emph{instances} of $\mathcal{P}$;
    \item $S$ is the set of \emph{solutions} of $\mathcal{P}$; and
    \item $R \subseteq \Lambda \times S$ is a binary relation that associates each instance $I \in \Lambda$ with a set $SOL(I) = \{s \in S: (I,s) \in R\}$. The elements in $SOL(I)$ are the  \emph{feasible solutions} of $\mathcal P$ for $I$.
\end{itemize}
A \emph{combinatorial optimization problem} $\mathcal{P}$ is a quintuple $\langle\Lambda, S, R, f_{\mathcal P}, g\rangle$, where:
\begin{itemize}
    \item $\langle\Lambda, S, R\rangle$ is a combinatorial search problem,
    \item $f_{opt}: S \rightarrow Y$ is the \emph{optimization function}, where $Y$ is a totally ordered set (usually $Y \in \{\mathbb{N}, \mathbb{R}\}$); 
    \item $g: 2^S \rightarrow S$ is the \emph{comparator function}, with $g \in \{\min, \max\}$.
\end{itemize}
An \emph{optimal solution} of $\mathcal P$ for $I$ is any feasible solution $s^*$ in $SOL(I)$ such that \linebreak
$s^* = \arg\min_{s \in SOL(I)} f_{opt}(s)$, if $g = \min$, and $s^* = \arg\max_{s \in SOL(I)} f_{opt}(s)$, if $g = \max$.
For ease of notation, in the following, we denote by $\solution(I)$, both a feasible solution of a combinatorial search problem and an optimal solution of a combinatorial optimization problem. Also, we refer to combinatorial search/optimization problems simply as \emph{combinatorial problems}.

\noindent
{\bf Dynamic programming.} 
Let $\mathcal P$ be a combinatorial problem and let $I$ be an instance of~$\mathcal P$. 
A (bottom-up) dynamic programming algorithm for $\mathcal P$ can be implemented by executing the following~steps:

\dfn{Dynamic Programming Steps:}{
\begin{description}
\item[\bf Table Setup:] Determine integers $d_1,d_2,\dots,d_k$ that depend on $I$ and create a table $D$ of dimension $d_1 \times d_2 \times \dots \times d_k$. Initialize some ``easy'' entries of $D$ ({\sc Base Case}) directly using information from $I$ and initialize the remaining ``difficult'' entries  with a default value representing the fact that such entries have not yet acquired their final value.
\item[\bf Table Update:] Compute the ``difficult'' entries of $D$ ({\sc Recursive Case}) by evaluating a {\bf recursive formula} that expresses each entry $D[i_1][i_2]\dots[i_k]$ in terms of a subset of the (already computed) entries $D[j_1][j_2 ]\dots[j_k]$ corresponding to optimal solutions for structurally related subproblems such that $(j_1,j_2\dots,j_k) \prec (i_1,i_2\dots,i_k)$.
\item[\bf Solution Retrieval:] Return the value contained in a specific entry of $D[i^*_1][i^*_2]\dots[i^*_k]$ (whose position $\langle i^*_1,i^*_2,\dots,i^*_k\rangle$ in $D$ depends on $I$ and $\mathcal P$) or retrieve a solution of $I$ by inspecting~$D$ (usually exploiting the information contained in $D[i^*_1][i^*_2]\dots[i^*_k]$).
\end{description}
}

Whereas the {\bf efficiency} of this algorithm design paradigm lies in the fact that solutions of {\em overlapping problems} are stored in the table and hence need to be computed only once, its {\bf correctness} lies in the fact that  ``difficult'' entries can be computed by exploiting the values of previously computed (``easy'' and ``difficult'') entries (by the {\em optimal substructure property}). Clearly, the {\bf Table Update} is the most interesting and challenging step in the overall approach. 
Fortunately, many optimization problems $\mathcal P$, including those considered in this paper, naturally exhibit a simple recursive formulation for the entries of their dynamic programming table $D$.
Consider an entry $D[i_1][i_2]\dots[i_k]$ of $D$.
Let $S_{i_1,i_2,\dots,i_k}$  be the \emph{dependency set} of $D[i_1][i_2]\dots[i_k]$, composed of the indices of the entries of $D$ on which the computation of the entry $D[i_1][i_2]\dots[i_k]$ depends. Observe that $S_{i_1,i_2,\dots,i_k}$ is some subset of $\mathbb{N}^k$ such that 
for each entry $\langle j_1,j_2,\dots,j_k\rangle$ in $S_{i_1,i_2,\dots,i_k}$ we have that  $\langle j_1,j_2,\dots,j_k\rangle \prec \langle i_1,i_2,\dots,i_k \rangle$.
Then, the recursive formula for $D[i_1][i_2]\dots[i_k]$ is of the~form:

\begin{equation}
\centering
\hfil
D[i_1][i_2]\dots[i_k] = \operator_{\mathcal{X} \in \mathcal C_{i_1,i_2,\dots,i_k}} f_{\mathcal P}(i_1,i_2,\dots,i_k,\mathcal{X}),\label{eq:dynamic-recurrence-generic}
\end{equation}

\noindent
where:
\begin{enumerate*}[label=({\bf \arabic*})]
\item $\operator \in \{\min,\max,\find,\findAll\}$;
\item $C_{i_1,i_2,\dots,i_k}$ is a set, called \emph{generating set},
composed of $h$-element subsets of $S_{i_1,i_2,\dots,i_k}$, where each subset provides the input to construct a particular candidate value for $D[i_1][i_2]\dots[i_k]$ (where $h$ is an integer constant, called \emph{dependency index}, determined by $\mathcal P$, 
which specifies the number of subinstances into which each instance is divided in the recursive definition); 
\item $f_{\mathcal P}$ is a function specific for problem $\mathcal P$ that computes a candidate value for $D[i_1][i_2]\dots[i_k]$,
assuming that all entries of $D$ with indices in $S_{i_1,i_2,\dots,i_k}$ have already been computed.
\end{enumerate*}

\begin{remark}
The {\em optimal substructure property} of a problem $\mathcal P$ is formally captured by the dependency set $S_{i_1,i_2,\dots,i_k}$, whose entries $\langle j_1,j_2,\dots,j_k\rangle$ must satisfy 
$\langle j_1,j_2,\dots,j_k\rangle \prec \langle i_1,i_2,\dots,i_k\rangle$.
\end{remark}

In most problems, the {dependency index} is a small integer, usually equal to 1 or 2.
A textbook example of a problem fitting \cref{eq:dynamic-recurrence-generic} with $h=1$ is the {\sc Coin Change} problem~\cite{DBLP:books/daglib/0015106}.
Given a set of positive integer coin denominations $c_1 < c_2 < \dots < c_r$ and a target sum $T$, the goal is to achieve $T$ using the fewest possible number of coins, assuming an unlimited supply of each denomination, or determine if it is not possible to obtain $T$. Let $D$ be a dynamic programming table of size $T+1$, whose entries $D[i]$ represent the minimum number of coins needed to make up the sum $i$, with $D[i]=\infty$ if it is not possible. The base case of the dynamic programming approach is $D[0]=0$. For the recursive case, to compute $D[i]$ for $i>0$, we have to consider all possible choices for the first coin.  Once the first coin is selected, the remaining amount must be reached optimally.
This can be expresses by the recursive formula
\begin{equation}
    \label{eq:conin-change}
    D[i]=
    \begin{cases}
        0, & \textit{if } i=0 \\
        \displaystyle\min_{j: c_j\leq i} (1+D[i-c_j]), & \textit{if } i>0
    \end{cases}
\end{equation}

\noindent Clearly, the recursive case of \cref{eq:conin-change} matches the pattern of \cref{eq:dynamic-recurrence-generic}. In fact, we can set $S_{i} = \{j: c_j \leq i\}$, $C_{i} = \{\{j\}: c_j \leq i\}$, $h=1$, and $f_{\mathcal P}(i,\{j\})=1+D[i-c_j]$.

A notable example of a problem fitting \cref{eq:dynamic-recurrence-generic} with $h=2$ is the {\sc Matrix Chain Multiplication} problem. Given a sequence of $n$ matrices, $A_1,A_2,\dots,A_n$, and their dimensions $p_0,p_1,p_2,\dots,p_n$, where for $i=1,2,\dots,n$, matrix $A_i$ has dimension $p_{i-1}\times p_i$, the problem asks to determine the order of matrix multiplications that minimizes the total number of scalar multiplications needed to obtain $A_1 \times A_{2} \times \dots \times A_n$. Note that, in the {\sc Matrix Chain Multiplication} problem, we are not actually multiplying matrices; instead, the goal is only to determine an order for multiplying matrices that has the lowest cost. 
Recall that matrix multiplication is associative, therefore we aim at grouping the above
multiplications to minimize the total number of scalar multiplications.

Let $D$ be a dynamic programming table of size $n \times n$, whose entries $D[i][j]$, with $1 \leq i \leq j \leq n$, store the minimum number of scalar multiplications needed to compute the product $A_i \times A_{i+1} \times \dots \times A_j$.
The base case of the dynamic programming approach is $D[i][i]=0$. For the recursive case, to compute $D[i][j]$ for $j>i$, we have to consider all possible choices to split the product at a matrix $A_k$, with $i \leq k < j$. 
This can be expressed by the recursive formula:

\begin{equation}
    \label{eq:matrix-chain-multiplication}
    D[i][j]=
    \begin{cases}
        0,& \textit{if } i=j\\
        \displaystyle\min_{i\leq k < j} (D[i][k] + D[k+1][j] + p_{i-1}p_kp_j), &\textit{if } i<j
    \end{cases}
\end{equation}

\noindent Clearly, the recursive case of \cref{eq:matrix-chain-multiplication} matches the pattern of \cref{eq:dynamic-recurrence-generic}. In fact, we can set 
$S_{i,j} = \{(i,k),(k+1,j): i \leq k < j\}$, 
$C_{i,j} = \{\{(i,k),(k+1,j)\}: i \leq k < j\}$, $h=2$, and $f_{\mathcal P}(i,j,\{(i,k),(k+1,j)\})=D[i][k] + D[k+1][j] + p_{i-1}p_kp_j$.

\section{Quantum Tools}\label{se:quantum-subroutines}
In this section, we provide the reader with the quantum primitives needed in this research. For a comprehensive introduction to the quantum computing field see, e.g., Nielsen and Chuang~\cite{DBLP:books/daglib/0046438}, and Aaronson~\cite{GroverAaronson}.

\emph{Qubits} are the fundamental units of quantum computing. They differ from classical bits in that they can exist in a superposition of the two classical states $\mathbf{0}$ and~$\mathbf{1}$. This unique characteristic enables quantum computers to perform, in parallel, multiple computations, allowing in some cases to achieve a substantial (and sometimes exponential) speedup over classical systems. Mathematically, a qubit is represented in a Hilbert space as a two-dimensional vector\footnote{In Dirac’s notation, a \emph{ket} such as $\ket{v}$, where $v$ is an arbitrary label, represents a vector corresponding to a quantum state.}
$\ket{\psi}= \binom{\alpha}{\beta} \in \mathbb{C}^2$, that is, $\ket{\psi} = \alpha \ket{0} + \beta \ket{1}$, where $\ket{0} = \binom{1}{0}$ and $\ket{1}= \binom{0}{1}$ are the two orthonormal basis states and the complex coefficients $\alpha$ and $\beta$ are the \emph{amplitudes} of such states.
The likelihood of measuring the qubit in state $\ket{0}$ or $\ket{1}$ is given by $|\alpha|^2$ and $|\beta|^2$, respectively.
Therefore, $\alpha$ and $\beta$ must satisfy the {\bf normalization condition} $|\alpha|^2 + |\beta|^2 = 1$, ensuring that the total probability remains $1$. 
A \emph{quantum state} over $n$ qubits is a unit vector in the Hilbert space $\mathbb{C}^{2^n}$.
The \emph{computational basis} $\{\ket{j}\}_{j \in [2^n]}$ consists of quantum states, where $|j\rangle$ is the unit vector with a $1$ in the $j$-th index and $0$ elsewhere.
A computational basis state $\ket{j}$ can be interpreted as a classical bit string $j$, allowing quantum systems to simulate classical algorithms. Any quantum state can be expressed as a weighted sum (or \emph{superposition}) of these basis states:

$$
    \ket{\Psi} = \sum_{j=0}^{2^n-1} \alpha_j \ket{j},
$$

\noindent 
where the amplitudes satisfy the normalization condition $\sum_{j=0}^{2^n-1} |\alpha_j|^2 = 1$.
If at least two coefficients $\alpha_j$ in the above expression are nonzero, the state is said to be in \emph{superposition}. 
When measuring $\ket{\Psi}$, the state collapses to $|j\rangle$ with probability $|\alpha_j|^2$.

In this paper, we focus on quantum computations performed in the {\bf circuit model of computation}. In this model, quantum algorithms are specified by \emph{quantum circuits}, obtained by composing quantum gates, that perform specific quantum computations. The process begins with the initialization of qubits, followed by the application of gate operations that modify their states. 
Finally, the circuit’s output is obtained by measuring the qubits. This operation collapses their quantum states into classical binary outcomes.
\emph{Quantum gates} are the fundamental building blocks of quantum circuits, analogous to classical logic gates in traditional computing. They are used to
manipulate quantum states while preserving key quantum properties like superposition and entanglement. 
Specifically, a quantum gate performs a linear transformation on its input quantum state, meaning that a superposition of states is mapped to the corresponding superposition of their images. In particular, any such a transformation $U$ must be \emph{unitary}, satisfying the condition $\mathbb{I} = U^\dag U = U U^\dag$, where $\mathbb{I}$ denotes the identity matrix and $U^\dag$ denotes the transpose conjugate of $U$.

As a consequence, quantum computation is inherently \emph{reversible} as long as no measurement is performed. That is, given an output quantum state
$\ket{\phi}$, obtained by applying $U$ to an initial quantum state $\ket{\psi}$, the original state can be fully recovered by applying $U^{-1}=U^\dag$ to $\ket{\phi}$. 
An important quantum gate (often used in the initialization of qubits) is the Hadamard gate $H = \frac{1}{\sqrt{2}}\big(\begin{smallmatrix} 1 & 1\\ 1 & -1\end{smallmatrix}\big)$, which can be used a preliminary step to transform the $\ket{0}$ state into to a uniform state $\frac{1}{\sqrt{2}}(\ket{0} + \ket{1})$ of the two basis states.

\emph{Quantum Random Access Memory (QRAM)} is the ``quantum analog'' of conventional RAM, designed to store and access data in a quantum format. Like RAM, QRAM consists of three main components: an input (or address) register $a$, an output (or data) register $d$, and memory arrays. However, unlike traditional RAM, QRAM’s input and output registers are composed of qubits rather than classical bits, while the memory arrays can be either classical or quantum, depending on the application~\cite{PhysRevLett.100.160501}.  
A key feature of QRAM lies in its method of memory access. Instead of retrieving data from a single memory location at a time, QRAM leverages quantum superposition to access multiple memory locations simultaneously. Specifically, while a classical RAM uses $n$  bits to randomly access one of  $N = 2^n$  distinct memory cells, a QRAM uses $n$  qubits to address a quantum superposition of all $N$  memory cells simultaneously. When a quantum computer requires access to a superposition of memory cells, the address register\footnote{A set comprising multiple qubits is a \emph{register}. 
A quantum register $a$ with $n$ qubits $\ket{q_i}$, with $i \in [n]$, is denoted as a tensor product
$\ket{\psi}_a = \ket{q_0} \otimes \ket{q_1} \otimes \cdots \otimes \ket{q_{n-1}}$.} $a$ must hold a superposition of addresses, represented as  $\sum_j \alpha_j \ket{j}_a$. In response, the QRAM returns a corresponding superposition of data in the data register $d$, ensuring that the retrieved data remains correlated with the address register:

$$\sum_j \alpha_j \ket{j}_a\ket{0^k}_d \overset{QRAM}{\longrightarrow} \sum_j \alpha_j \ket{j}_a \ket{\delta_j}_d,$$
where  $\delta_j$  represents the content of the  $j$-th memory cell, $k$ bits suffice to classically encode the content of any memory cell, and $\ket{0^k}$ denotes the quantum basis state composed of $k$ qubits set to $\ket{0}$. QRAM plays a crucial role in several quantum algorithms, such as quantum searching~\cite{DBLP:conf/stoc/Grover96}, minimum/maximum finding~\cite{DBLP:journals/corr/quant-ph-9607014}, counting~\cite{DBLP:conf/icalp/BrassardHT98}, period finding and discrete logarithm~\cite{DBLP:journals/siamrev/Shor99}, to cite a few. 
In particular, the use of QRAM enables us to exploit quantum search primitives that involve condition checking on data stored in random access memory. Specifically, the QRAM may be used by an oracle to check conditions based on the data stored in memory, marking the superposition states that correspond to feasible or optimal solutions \cite{PhysRevLett.100.160501, DBLP:journals/sensors/PhalakCG23, DBLP:conf/micro/XuHFG023, DBLP:journals/access/ZidanAKAE21}.

We now describe the {\bf quantum subroutines} used in our algorithms. Observe that every function $f$ implemented as a classical circuit can be approximated with arbitrary accuracy, using a discrete set of quantum gates, by a quantum circuit with only a polylogarithmic overhead~\cite{DBLP:journals/qic/DawsonN06,DBLP:books/daglib/0046442,DBLP:books/daglib/0046438}.
Therefore, in the following, we assume that every classical function $f$, provided as classical circuit $C_f$, can be passed to and accessed by our quantum procedures as a quantum version of $C_f$. In the remainder, for any computable function $f$, we use the notation $U_f$ to denote the corresponding quantum circuit.

Let $W$ be a ground set and let $\omega$ be an element of $W$. Let $bin(\omega)$ be the integer corresponding to a binary encoding of $\omega$. For ease of notation, in the remainder, we will denote the state $\ket{bin(\omega)}$ simply as $\ket{\omega}$. In particular, given an element ${\cal X} \in C_{i_1,i_2,\dots,i_k}$, we will use the state $\ket{{\cal X}}$.

Consider a computational problem $\mathcal P$ that can be solved via dynamic programming building a $k$-dimensional table $D$ of size $d_1 \times \dots \times d_k$. Recall that by $C_{i_1,i_2,\dots,i_k}$ we denote the generating set of an entry $D[i_1][i_2]\dots[i_k]$ of $D$. We let $\lambda_{i_1,i_2,\dots,i_k} = |C_{i_1,i_2,\dots,i_k}|$ and we let $\sigma = \sum^k_{i=1}\log{d_i}$.

\begin{algorithm}[tb!]
\begin{algorithmic}[1]
\Procedure{\texttt{StatePrep}}{$i_1 i_2 \dots i_k$}
\State \textbf{input:} Registers $\ket{i_1 i_2 \dots i_k}$ and an $\ell$-bit zero register $\ket{0^{\ell}}$
\State \textbf{output:} Uniform superposition state $\ket{\Psi_{i_1,i_2,\dots,i_k}}$
\State Apply the quantum circuit $U_{f_C}$ to determine $\lambda_{i_1,i_2,\dots,i_k} = |C_{i_1,i_2,\dots,i_k}|$
\State Prepare the state $\ket{i_1 i_2 \dots i_k}\ket{\lambda_{i_1,i_2,\dots,i_k}}$
\State Introduce an additional $\log(\lambda_{i_1,i_2,\dots,i_k})$-bit zero register $\ket{0^{\log \lambda_{i_1,i_2,\dots,i_k}}}$
\State Apply Hadamard gates to the qubits of the last register to prepare the uniform superposition: $$\ket{\Psi_{i_1,i_2,\dots,i_k}} = \frac{1}{\sqrt{\lambda_{i_1,i_2,\dots,i_k}}}\ket{i_1,i_2,\dots,i_k}\ket{\lambda_{i_1,i_2,\dots,i_k}}\sum^{\lambda_{i_1,i_2,\dots,i_k}-1}_{u=0}\ket{u}$$
\State \Return $\ket{\Psi_{i_1,i_2,\dots,i_k}}$
\EndProcedure
\end{algorithmic}
\caption{Procedure {\sc StatePrep} prepares a uniform superposition over the set $C_{i_1,i_2,\dots,i_k}$.}
\label{algo:preparation}
\end{algorithm}

\noindent{\sc \underline{Subroutine StatePrep}}. The first subroutine prepares a uniform superposition $\Psi_{i_1,i_2,\dots,i_k}$ of the integers in $[\lambda_{i_1,i_2,\dots,i_k}]$; see the pseudocode of \cref{algo:preparation}. It assumes the existence of a classical procedure $f_C$ that determines $\lambda_{i_1,i_2,\dots,i_k}$ given the values $i_1,i_2,\dots,i_k$. The subroutine starts with the register $\ket{i_1 i_2 \dots i_k}$ and with a register storing an $\ell$-bit zero string $\ket{0^{\ell}}$ (where the value of $\ell$ will be discussed later). It exploits the quantum gate $U_{\lambda}$ to determine $\lambda_{i_1,i_2,\dots,i_k}$ and prepares the state $\ket{i_1 i_2 \dots i_k}\ket{\lambda_{i_1,i_2,\dots,i_k}}$. 
Then, the subroutine takes in input an additional register storing a $\log(\lambda_{i_1,i_2,\dots,i_k})$-bit zero string and applies Hadamard gates to each of the qubits of such a register to prepare the uniform superposition state:
$$\ket{\Psi_{i_1,i_2,\dots,i_k}} = \frac{1}{\sqrt{\lambda_{i_1,i_2,\dots,i_k}}}\ket{i_1 i_2 \dots i_k}\ket{\lambda_{i_1,i_2,\dots,i_k}}\sum^{\lambda_{i_1,i_2,\dots,i_k}-1}_{u=0}\ket{u}$$
We use the signature \texttt{StatePrep($i_1,i_2,\dots,i_k$)} to denote calls to the subroutine {\sc StatePrep}.

\begin{algorithm}[tb!]
\begin{algorithmic}[1]
\Procedure{$\texttt{TableIndexPrep}$}{$i_1,i_2,\dots,i_k,u$}
\State \textbf{input:} Registers $\ket{i_1,i_2,\dots,i_k}$, the register $\ket{u}$, and an $h\cdot \sigma$-bit zero register $\ket{0^{h\cdot \sigma}}$
\State \textbf{output:} State $\ket{E_{i_1,i_2,\dots,i_k,u}}$ encoding the $u$-th element $\gamma(\langle i_1,i_2,\dots,i_k\rangle, u)$ of $C_{i_1,i_2,\dots,i_k}$
\State Apply the quantum circuit $U_\gamma$ to retrieve the element $X_u = \gamma(\langle i_1,i_2,\dots,i_k\rangle, u)$ 
\State Store  $X_u$ in the output register, thus obtaining the state:
$$\ket{E_{i_1 i_2 \dots i_k,u}}=\ket{i_1 i_2 \dots i_k}\ket{u}\ket{
X_u}$$

\State \Return $\ket{E_{i_1 i_2 \dots i_k,u}}$
\EndProcedure
\end{algorithmic}
\caption{Procedure {\sc TableIndexPrep} for retrieving the $u$-th element of $C_{i_1,i_2,\dots,i_k}$.}
\label{algo:candidate}
\end{algorithm}

\noindent
\noindent{\sc \underline{Subroutine TableIndexPrep}}. 
The second quantum subroutine prepares the state \linebreak $\ket{E_{i_1 i_2 \dots i_k,u}}$ that correlates the integers in $[\lambda_{i_1,i_2,\dots,i_k}]$ with the elements of $C_{i_1 i_2 \dots i_k}$; 
see the pseudocode of \cref{algo:candidate}.
Our quantum subroutine assumes the existence of a classical {\bf injective function} $\gamma: \mathbb{N}^{k} \times \mathbb{N} \rightarrow C_{i_1 i_2 \dots i_k}$ that maps the pair 
$\langle\langle i_1 i_2 \dots i_k \rangle, u\rangle$ to 
an element of $C_{i_1 i_2 \dots i_k}$. 
Observe that the elements of $C_{i_1,i_2,\dots,i_k}$ admit a binary representation $bin(\mathcal X)$ of length $h \cdot \sigma$.
The subroutine starts with 
the register $\ket{i_1 i_2 \dots i_k}$, the register $\ket{u}$, and with a register storing an $(h \cdot \sigma)$-bit zero string $\ket{0^{h \cdot \sigma}}$ and uses $U_\gamma$ to prepare the state:  
$$\ket{E_{i_1 i_2 \dots i_k,u}}=\ket{i_1 i_2 \dots i_k}\ket{u}\ket{\gamma(\langle i_1,i_2,\dots,i_k\rangle, u)}.$$
We use the signature \texttt{TableIndexPrep($i_1,i_2,\dots,i_k,u$)} to denote calls to the subroutine {\sc TableIndexPrep}.

\noindent{\sc \underline{Subroutine DP}}. The third quantum subroutine uses either 
\begin{enumerate*}[label=({\bf \roman*})]
\item the  quantum min/max finding algorithms ({\sc QMin} and {\sc QMax}) due to Dürr and Høyer~\cite{DBLP:journals/corr/quant-ph-9607014}, or 
\item the quantum finding algorithm ({\sc QFind}) due to Grover to search for an item satisfying a condition in an unsorted list~\cite{GroverAaronson}
or 
\item the quantum finding algorithm ({\sc QFindAll}) due to Ambainis to search for all items satisfying a condition in an unsorted list~\cite{10.1145/992287.992296}. \end{enumerate*}
In particular, {\sc QMin} and {\sc QMax} allow to determine the element in a ground set of size $N$ that minimizes/maximizes a given function in $\tilde{\bigO}(\sqrt{N})$ time, {\sc QFind} (resp.\ {\sc QFindAll}) allows to determine an element (resp. all the elements) in a ground set of size $N$ that satisfy a certain condition in $\tilde{\bigO}(\sqrt{N})$ time (resp. $\sqrt{NM}$ time, where $M$ is the number of elements satisfying the condition).

\begin{algorithm}[tb!]
\begin{algorithmic}[1]
\Procedure{$\texttt{DP}$}{$i_1,i_2,\dots,i_k, f_{\mathcal P}, op,  \lessdot =  \texttt{null}$}
\State \parbox[t]{\dimexpr\linewidth-\algorithmicindent}{\textbf{input:} Registers $\ket{i_1,i_2,\dots,i_k}$, the function $f_{\mathcal P}$, and an operator $\operator \in \{\min, \max, \find, \findAll\}$. Additionally, a comparator $\lessdot$, if $\operator \in \{\min, \max\}$. \strut}
\State \parbox[t]{\dimexpr\linewidth-\algorithmicindent}{\textbf{output:} The value $\operator_{\mathcal{X} \in \mathcal C_{i_1,i_2,\dots,i_k}} f_{\mathcal P}(i_1,i_2,\dots,i_k,\mathcal{X})$ to be assigned to $D[i_1][i_2]\dots [i_k]$ \strut}
\State \parbox[t]{\dimexpr\linewidth-\algorithmicindent}{Invoke the subroutine \texttt{StatePrep}($i_1,i_2,\dots,i_k$) to prepare the superposition state $\ket{\Psi_{i_1,i_2,\dots,i_k}}$ \strut}
\State \parbox[t]{\dimexpr\linewidth-\algorithmicindent}{Apply the subroutine \texttt{TableIndexPrep}($i_1,i_2,\dots,i_k,u$), in parallel to each basis state of $\ket{\Psi_{i_1,i_2,\dots,i_k}}$, to prepare the superposition $\ket{\Phi_{i_1,i_2,\dots,i_k}}$ \strut}
\If {$\operator = \min$}
\State Apply \texttt{QMaxMin}($i_1,i_2,\dots,i_k, f_{\mathcal P}, \lessdot$) to $\ket{\Phi_{i_1,i_2,\dots,i_k}}$
\ElsIf {$\operator = \max$}
\State Apply \texttt{QMaxMax}($i_1,i_2,\dots,i_k, f_{\mathcal P}, \lessdot$) to $\ket{\Phi_{i_1,i_2,\dots,i_k}}$
\ElsIf {$\operator$ is $\find$}
\State Apply \texttt{QFind}($i_1,i_2,\dots,i_k, f_{\mathcal P}$) to $\ket{\Phi_{i_1,i_2,\dots,i_k}}$
\ElsIf {$\operator$ is $\findAll$}
\State Apply \texttt{QFindAll}($i_1,i_2,\dots,i_k, f_{\mathcal P}$) to $\ket{\Phi_{i_1,i_2,\dots,i_k}}$
\EndIf
\State \Return $\operator_{\mathcal{X} \in \mathcal C_{i_1,i_2,\dots,i_k}} f_{\mathcal P}(i_1,i_2,\dots,i_k,\mathcal{X})$
\EndProcedure
\end{algorithmic}
\caption{Procedure {\sc DP} for computing the value $\operator_{\mathcal{X} \in \mathcal C_{i_1,i_2,\dots,i_k}} f_{\mathcal P}(i_1,i_2,\dots,i_k,\mathcal{X})$. We use the signature 
\texttt{QMaxMin}($i_1,i_2,\dots,i_k, f_{\mathcal P}, \lessdot$),  \texttt{QMaxMax}($i_1,i_2,\dots,i_k, f_{\mathcal P}, \max, \lessdot$), 
\texttt{QFind}($i_1,i_2,\dots,i_k, f_{\mathcal P}$), and  
\texttt{QFindAll}($i_1,i_2,\dots,i_k, f_{\mathcal P}$) to denote calls to the algorithms {\sc QMaxMin}, {\sc QMaxMin}, {\sc QFind}, and {\sc QFindAll}, respectively.}
\label{algo:dynprogalgo}
\end{algorithm}

The subroutine (see the pseudocode of \cref{algo:dynprogalgo}) takes as input at least:
\begin{enumerate*}[label=({\bf \arabic*})]
\item 
the integers $i_1, i_2, \dots, i_k$;
\item 
the function $f_{\mathcal P}$; and
\item an operator $\operator \in \{\min, \max, \find, $ $ \findAll\}$.
\end{enumerate*}
Furthermore, if $\operator \in \{\min, \max\}$, it additionally takes as an input a comparator,~$\lessdot$, to maximize (or minimize) over, that defines a total ordering of the values in the codomain of $f_{\mathcal P}$ (i.e., the values stored in $D$).
First, the subroutine invokes \texttt{StatePrep} $(i_1, i_2, \dots, i_k)$ to prepare the superposition state $\ket{\Psi_{i_1,i_2,\dots,i_k}}$
and then applies \texttt{TableIndexPrep}, in parallel to each basis state of $\ket{\Psi_{i_1,i_2,\dots,i_k}}$, to prepare the superposition state: %

$$
\ket{\Phi_{i_1,i_2,\dots,i_k}} = \frac{1}{\sqrt{\lambda_{i_1,i_2,\dots,i_k}}}\ket{i_1,i_2,\dots,i_k}\ket{\lambda_{i_1,i_2,\dots,i_k}}\sum^{\lambda_{i_1,i_2,\dots,i_k}-1}_{u=0}\ket{u}\ket{\gamma(\langle i_1,i_2,\dots,i_k\rangle, u)}.$$

\begin{remark}
The time $T'_{i_1,i_2,\dots,i_k}$ required to prepare the state $\ket{\Phi_{i_1,i_2,\dots,i_k}}$ is bounded by the time needed to execute gates $U_{f_C}$ and $U_\gamma$.
\end{remark}

Observe that each of the states $\ket{i_1,i_2,\dots,i_k}\ket{\lambda_{i_1,i_2,\dots,i_k}}\ket{u}\ket{\gamma(\langle i_1,i_2,\dots,i_k\rangle, u)}$ composing $\ket{\Phi_{i_1,i_2,\dots,i_k}}$ provides the input for computing, using $f_{\mathcal P}$ and the quantum search subroutines described above, a candidate value for $D[i_1][i_2]\dots[i_k]$, that is, the registers $\ket{i_1,i_2,\dots,i_k}$ and $\ket{\gamma(\langle i_1,i_2,\dots,i_k\rangle, u)}$.
Therefore, the subroutine proceeds by applying the algorithm {\sc QMin}, {\sc QMax}, {\sc QFind}, or {\sc QFindAll}, depending on whether the chosen operator $\operator$ is equal to $\min$ (or $\max$), $\find$, or $\findAll$, respectively, 
{\bf only to} the candidate values for $D[i_1][i_2]\dots[i_k]$ determined by the entries in $C_{i_1,i_2,\dots,i_k}$.

We use the signatures \texttt{DP($i_1,i_2,\dots,i_k, f_{\mathcal P}, op$)} and
\texttt{DP($i_1,i_2,\dots,i_k, f_{\mathcal P}, op, \lessdot$)} to denote calls to the subroutine {\sc DP}, when $\operator \in \{\find,\findAll\}$ or $\operator \in \{\min,\max\}$, respectively.

Altogether, we obtain the following main algorithmic theorem.

\begin{theorem}\label{th:minmax}
Let $\mathcal P$ be a combinatorial problem, let $I$ be an instance of $\mathcal P$, and let $n$ be the 
size of $I$. 
Also, let
$T_{\mathcal P}$ be the time needed to compute quantumly the function~$f_{\mathcal P}$ and let $T'_{i_1,i_2,\dots,i_k}$ be the time needed to prepare $\ket{\Phi_{i_1,i_2,\dots,i_k}}$.
Suppose that each of the values of the entries of $D$ can be represented using $w$ bits with $w \in \bigO(polylog(n))$. 
 Then, the following holds for subroutine \texttt{DP}:
\begin{itemize}
\item If $\operator = \find$, \texttt{DP} determines a value of $f_{\mathcal P}$ for an entry $D[i_1][i_2]\dots[i_k]$ in $\tilde{\bigO}(T'_{i_1,i_2,\dots,i_k} + T_{\mathcal P}\sqrt{\lambda_{i_1,i_2,\dots,i_k}})$ time. 
\item If $\operator = \findAll$, \texttt{DP} determines all the $M$ possible distinct values of $f_{\mathcal P}$ for an entry $D[i_1][i_2]\dots[i_k]$ in $\tildeO(T'_{i_1,i_2,\dots,i_k} + T_{\mathcal P}\sqrt{\lambda_{i_1,i_2,\dots,i_k}M})$ time. 
\item If $\operator \in \{\min,\max\}$, 
consider some total ordering defined by $\lessdot$ of the data values in $D$ such that comparison according to such an ordering can be performed in $\bigO(w)$ time. Then, \texttt{DP} determines the minimum (or maximum) value of $f_{\mathcal P}$ for an entry $D[i_1][i_2]\dots[i_k]$, under the specified ordering, in $\tildeO(T'_{i_1,i_2,\dots,i_k} + T_{\mathcal P}\sqrt{\lambda_{i_1,i_2,\dots,i_k}})$ time. 
\end{itemize}
\end{theorem}

\section{Quantum Dynamic Programming}

In this section, we describe a framework that exploits the quantum subroutines defined in \cref{se:quantum-subroutines} to 
obtain quantum speedups for many computational problems solved classically using dynamic programming algorithms. 

The recursive formula that specifies how to compute the entries of the dynamic programming table $D$ used for solving a computational problem $\mathcal P$ on a instance $I$ determines a {\bf dependency digraph}  $\depgr$. The nodes of $\depgr$ are in one-to-one correspondence with the entries of $D$, i.e., the subproblems of $\mathcal P$ defined by $I$. For each node
$n_{i_1,i_2,\dots,i_k}$ of $\depgr$ associated with the entry $D[i_1][i_2]\dots[i_k]$ of $D$, graph  $\depgr$ contains an arc directed from $n_{i_1,i_2,\dots,i_k}$ to each of the nodes $n_{j_1,j_2,\dots,j_k}$ associated with the entries $D[j_1][j_2]\dots[j_k]$ that occur in the recursive formula for $D[i_1][i_2]\dots[i_k]$.
These are the entries indexed by the tuples $\langle j_1, j_2,\dots,j_k \rangle \in S_{i_1,i_2,\dots,i_k}$.
Clearly, by the optimal substructure property, $\depgr$ is a directed acyclic graph.

Let $\mathcal P$ be a combinatorial problem and suppose that it admits a dynamic programming algorithm that relates the dependency set $S_{i_1,i_2,\dots,i_k}$ and the generating set $C_{i_1,i_2,\dots,i_k}$ of an entry $D[i_1][i_2]\dots[i_k]$ as follows. For each element $\langle j_1,j_2,\dots,j_k\rangle \in S_{i_1,i_2,\dots,i_k}$, there exists a unique set ${\cal X} \in C_{i_1,i_2,\dots,i_k}$ such that $\langle j_1,j_2,\dots,j_k\rangle  \in {\cal X}$.  
We say that a dynamic programming algorithm exhibiting the above characteristic  is \emph{simple} and that $\mathcal P$ is a \emph{simple problem}. 
Observe that, for a simple dynamic programming algorithm, it holds that $\lambda_{i_1,i_2,\dots,i_k} = |C_{i_1,i_2,\dots,i_k}| = \frac{|S_{i_1,i_2,\dots,i_k}|}{h}$. 
In particular, this implies that, for each node $n_{i_1,i_2,\dots,i_k}$ of $\depgr$, it holds that $ \deg_{out}(n_{i_1,i_2,\dots,i_k}) = |S_{i_1,i_2,\dots,i_k}| = h \cdot \lambda_{i_1,i_2,\dots,i_k}$.
We have that several combinatorial problems, including those listed in \cref{tab:complexities}, as we prove later, are simple problems.

The next lemma shows how classical simple dynamic programming algorithms can be quantumly enhanced to reduce their time complexity, while maintaining the same storage requirements as in the classical setting. Such an improvement applies to problems whose recursive formulation satisfies \cref{eq:dynamic-recurrence-generic}.
\cref{tab:complexities} provides an overview of the speedups obtainable via \cref{thm:quantum-dp-lemma} for many well-known problems; see also \cref{se:appendix-uniform}.

\begin{theorem}\label{thm:quantum-dp-lemma}
Let $\mathcal P$ be a simple combinatorial problem, let $I$ be an instance of $\mathcal P$, and let $n$ be the size of $I$.
 Suppose that each of the values of the entries of $D$ can be represented using $w$ bits with $w \in \bigO(polylog(n))$.
Consider a simple classical dynamic programming algorithm $\mathcal A$ that solves $\mathcal P$ for $I$ by computing each entry $D[i_i][i_2]\dots[i_k]$ of a table $D$ of size $d_1 \times d_2 \times \dots \times d_k$, where the values $d_i$ depend on $I$, using a recurrence formula of the same form as \cref{eq:dynamic-recurrence-generic}:

$$D[i_1][i_2]\dots[i_k] = \operator_{\mathcal{X} \in \mathcal C_{i_1,i_2,\dots,i_k}} f_{\mathcal P}(i_1,i_2,\dots,i_k,\mathcal{X}),
$$

\noindent
where $\operator \in \{\min,\max,\find,\findAll\}$.
Also, let
$T_{\mathcal P}$,  $T_{f_C}$ and $T_\gamma$ be the time needed to compute quantumly the functions $f_{\mathcal P}$, $f_C$, and $\gamma$, respectively. 
Finally, let $T' = T_{f_C} + T_\gamma$, let $M$ be the maximum number of solutions of any subproblem of $\mathcal P$ for $I$, and let $\delta$ be the average degree of $\depgr$.

Then, there exists a quantum dynamic programming algorithm $Q_{\mathcal{A}}$ that solves $\mathcal P$ for $I$, using QRAM, with the following time and space bounds:

\begin{itemize}
\item If $\operator \in \{\find,\min,\max\}$, then
$Q_{\mathcal A}$ solves $\mathcal P$ for $I$ using
$\tildeO{\big(|V(\depgr)|(T' + T_{\mathcal P}\sqrt{\delta})\big)}$ time and $\tildeO{(|V(\depgr)|)}$ space.
\item If $\operator = \findAll$, then
$Q_{\mathcal A}$ solves $\mathcal P$ for $I$ using
$\tildeO{\big(|V(\depgr)|(T' + T_{\mathcal P}\sqrt{\delta \cdot M})\big)}$ time and $\tildeO{\big(M|V(\depgr)|\big)}$ space.

\end{itemize}
\end{theorem}

\begin{proof} In the following, we describe the algorithm $Q_{\mathcal A}$ and prove its space and time complexity. 
We describe $Q_{\mathcal A}$ in terms of the three steps that implement the dynamic programming approach provided by $\mathcal A$.

\noindent
{\bf Table Setup:}
As in the classical scenario, the algorithm starts by initializing a dynamic programming table $QD$ of size $d_1 \times d_2 \times \dots \times d_k$. Differently from the classical dynamic programming table $D$ used by $\mathcal A$, however, the table $QD$ is stored in QRAM. 
Suppose that each of the values of the entries of $D$ (and thus of $QD$) can be represented using $w$ bits.
Let $f_{init}$ be the classical function that, provided with the tuple $\langle i_1,i_2,\dots,i_k\rangle$ that addresses the entry $D[i_1][i_2]\dots[i_k]$, computes the initialization value of $D[i_1][i_2]\dots[i_k]$.
To this aim, for each $\langle i_1,i_2,\dots,i_k\rangle \in [d_1]\times[d_2]\times \dots \times [d_k]$, we provide the gate $U_{init}$ with the address register  $\ket{i_1 i_2 \dots i_k}_a$ and with a data register storing a $w$-bit zero string $\ket{0^w}$. The application of $U_{init}$ to these registers results in the state $\ket{i_1 i_2 \dots i_k}_a \ket{f_{init}(\langle i_1,i_2,\dots,i_k\rangle)}_d$, which forms the input for a \texttt{QRAM write} operation that initializes $QD[i_1][i_2]\dots[i_k]$.

\noindent
{\bf Table Update:} We process the tuples $\langle i_1,i_2,\dots,i_k\rangle \in [d_1]\times[d_2]\times \dots \times [d_k]$ in lexicographic order.
For each tuple $\langle i_1,i_2,\dots,i_k\rangle$, the algorithm invokes either the quantum subroutine \texttt{DP($i_1,i_2,\dots,i_k, f_{\mathcal P}, op$)} or
\texttt{DP($i_1,i_2,\dots,i_k, f_{\mathcal P}, op, \lessdot$)}, based on whether $\operator \in \{\find,\findAll\}$ or $\operator \in \{\min,\max\}$, respectively. 
This allows us to compute $\operator_{\mathcal{X} \in \mathcal C_{i_1,i_2,\dots,i_k}} f_{\mathcal P}(i_1,i_2,\dots,i_k,\mathcal{X})$, which is stored in an output register $\ket{d_{i_1,i_2,\dots,i_k}}_o$ correlated with the address register $\ket{i_1 i_2 \dots i_k}_a$. Finally, the state 
$\ket{i_1 i_2 \dots i_k}_a\ket{d_{i_1,i_2,\dots,i_k}}_o$ forms the input for a \texttt{QRAM write} operation that updates $QD[i_1][i_2]\dots[i_k]$.

\noindent
{\bf Solution Retrieval:} Let $\langle i^*_1, i^*_2,\dots,i^*_k\rangle$ be the entry of $D$ whose value determines the solution to $\mathcal P$. Then, a solution for $\mathcal P$ can be obtained by reading the entry $QD[i^*_1][i^*_2]\dots[i^*_k]$.

 Next, we analyze the space complexity of $Q_{\mathcal A}$. 
 As in its classic counterpart, the space complexity of $Q_{\mathcal A}$ is asymptotically bounded by the storage requirement of the dynamic programming table, which can be upper bounded by multiplying the number of entries of $QD$, the number $w$ of bits to represent each of the values of the entries of $QD$ (i.e., the solutions to the subproblems of $\mathcal P$), and the maximum number $M$ of solutions of any subproblem of $\mathcal P$ for $I$. Also, recall that the entries of $QD$ are in one-to-one correspondence with the nodes of $\depgr$.
Therefore, the space complexity of $Q_{\mathcal A}$ is in ${\bigO}\big(w(|V(\depgr)|\big)$, if $\operator = \{ \find, \min, \max \}$, and 
in ${\bigO}\big((w \cdot M)|V(\depgr)|\big)$, if $\operator = \findAll$. 

Finally, we analyze the time complexity of $Q_{\mathcal A}$. Obviously, the time needed to compute function $f_{init}$ is in $O(T_{\mathcal P})$. Therefore, the time complexity $\Time(I)$ of $Q_{\mathcal A}$ on input $I$ is asymptotically bounded by the time required to execute the {\bf Table Update} step. We now give the corresponding bound. For simplicity, we assume first that $\operator  \in \{\find,\min,\max\}$.
Let $n = |V(\depgr)|$  and recall that $\delta$ denotes the average degree of $\depgr$. In order to upper bound $\Time(I)$, we proceed as follows. For $i = 1,\dots,{\log \log n}$, let $V_i$ be the subset of $V(\depgr)$ whose degree is between $2^{2^{i-1}} \delta$ and $2^{2^{i}} \delta$, and let $V_0$ be the subset of $V(\depgr)$ whose degree is less than $2\delta$. 
Recall that, by \cref{th:minmax}, the quantum subroutine \texttt{DP} determines a value of $f_{\mathcal P}$ for an entry $D[i_1][i_2]\dots[i_k]$ in $\tilde{\bigO}(T'_{i_1,i_2,\dots,i_k} + T_{\mathcal P}\sqrt{\lambda_{i_1,i_2,\dots,i_k}})$ time, that $T'_{i_1,i_2,\dots,i_k} \leq T'$, and that, for each node $n_{i_1,i_2,\dots,i_k}$ of $\depgr$, it holds that $\lambda_{i_1,i_2,\dots,i_k} \leq {\deg_{out}(n_{i_1,i_2,\dots,i_k})}$ since $\mathcal{P}$ is simple. Therefore, we~have:
\begin{equation}\label{eq:time-complexity}
\begin{split}
\Time(I) &= 
\sum_{n_{i_1,i_2,\dots,i_k} \in V(\depgr)} \tildeO\Big(T'_{i_1,i_2,\dots,i_k} + T_{\mathcal P}\sqrt{\lambda_{i_1,i_2,\dots,i_k}}\Big)\leq\\
&\leq 
\sum_{n_{i_1,i_2,\dots,i_k} \in V(\depgr)} \tildeO\Big(T' + T_{\mathcal P}\sqrt{\deg_{out}(n_{i_1,i_2,\dots,i_k})}\Big)\leq
\end{split}
\end{equation}
\begin{equation*}
\begin{split}
&\leq\tildeO\Big(|V_0|(T' + T_{\mathcal P}\sqrt{2\delta})\Big)+
\sum^{\log \log n}_{i=1} \tildeO\Big(|V_i| \big(T' + T_{\mathcal P}\sqrt{2^{2^i}\delta})\Big)
= \\
&= \tildeO(|V_0|T')+\tildeO(|V_0|T_{\mathcal P}\sqrt{\delta})+\sum^{{\log \log n}}_{i=1} \tildeO\big(|V_i| T'\big) + (T_{\mathcal P}\cdot \sqrt{\delta})\sum^{{\log \log n}}_{i=1} \tildeO\big(|V_i|2^{2^{i-1}}\big) 
\end{split}
\end{equation*}
\noindent Next, we provide an upper bound for~$|V_i|$.
Clearly, 
there exist at most $2n$ nodes $n_{i_1,i_2,\dots,i_k}$ of $\depgr$ such that $\deg_{out}(n_{i_1,i_2,\dots,i_k}) < 2\delta$ (that is, $|V_0| \leq 2n$).
Also, there exist at most $n$ nodes $n_{i_1,i_2,\dots,i_k}$ of $\depgr$ such that $2\delta \leq \deg_{out}(n_{i_1,i_2,\dots,i_k}) < 4 \delta$ (that is, $|V_1| \leq n$).
Similarly, there exist at most $\frac{n}{2}$ nodes $n_{i_1,i_2,\dots,i_k}$ of $\depgr$ such that $4\delta \leq \deg_{out}(n_{i_1,i_2,\dots,i_k}) < 16 \delta$ (that is, $|V_2| \leq \frac{n}{2}$).
More in general, for $i = 1, \dots, {\log \log n}$, there exist at most $\frac{n}{2^{2^{i-1}-1}}=\frac{2n}{2^{2^{i-1}}}$ nodes of $\depgr$ such that $2^{2^{i-1}} \delta \leq \deg_{out}(n_{i_1,i_2,\dots,i_k}) < 2^{^i} \delta$ (that is, $|V_i| \leq \frac{n}{2^{2^{i-1}-1}}=\frac{2n}{2^{2^{i-1}}}$). In \cref{eq:time-complexity}, we can then upper bound 
$V_0$ with $2n$, and  
$|V_i|$ with $\frac{2n}{2^{2^{i-1}}}$, if $i>1$. Therefore, we have:

\begin{equation}\label{eq:final}
\begin{split}
\Time(I) & =
 \tildeO(|V_0|T')+\tildeO(|V_0|T_{\mathcal P}\sqrt{\delta})+\sum^{{\log \log n}}_{i=1} \tildeO\big(|V_i| T'\big) + (T_{\mathcal P}\cdot \sqrt{\delta})\sum^{{\log \log n}}_{i=1} \tildeO\big(|V_i|2^{2^{i-1}}\big) \leq \\
&\leq \tildeO(n \cdot T'\big)+\tildeO(nT_{\mathcal P}\sqrt{\delta})+\tildeO(n \cdot T'\big)+ (T_{\mathcal P} \cdot \sqrt{\delta})\sum^{{\log \log n}}_{i=1} \tildeO\big(|V_i|2^{2^{i-1}}\big) \leq\\
& \leq  \tildeO(n \cdot T'\big)+\tildeO(nT_{\mathcal P}\sqrt{\delta})+  (T_{\mathcal P} \cdot \sqrt{\delta})\sum^{{\log \log n}}_{i=1} \tildeO\big(\frac{2n}{2^{2^{i-1}}}2^{2^{i-1}}\big) = \\
& =  \tildeO(n \cdot T'\big)+\tildeO(nT_{\mathcal P}\sqrt{\delta})+ (T_{\mathcal P} \cdot \sqrt{\delta}) \tildeO(n)\sum^{{\log \log n}}_{i=1}2 \\
&= 
\tildeO\Big(n \big( T' +  (T_{\mathcal P} \cdot \sqrt{\delta})(1 + 2{\log \log n})
\big)\Big)
= \tildeO\Big(n \big (T' + T_{\mathcal P} \cdot \sqrt{\delta})\Big)
\end{split}
\end{equation}

Altogether, \cref{eq:final} shows the desired bound for the running time of $Q_{\mathcal A}$ when \linebreak $\operator  \in \{\find,\min,\max\}$. For the case when $\operator = \findAll$ the same analysis applies with $\sqrt{\delta}$ replaced by $\sqrt{\delta \cdot M}$ according to \cref{th:minmax}. This concludes the proof.
\end{proof}

\subsection{Quantum Bellman-Ford for Single-Source Shortest Paths}\label{sc:Bellman-Ford}

In this section, we consider the {\sc Single-Source Shortest Paths} (SSSP) problem. Since the celebrated Bellman-Ford algorithm for SSSP is based on a dynamic programming recurrence that respects \cref{eq:dynamic-recurrence-generic}, it provides a primary example for a computational problem amenable to a quantum speedup via  \cref{thm:quantum-dp-lemma}.

In the following for the SSSP problem and in \cref{se:appendix-uniform} for the remaining problems listed in \cref{tab:complexities}, we demonstrate the applicability of \cref{thm:quantum-dp-lemma} as follows. 
First, we present the corresponding classical dynamic programming algorithm. In particular, we describe (i) the {\bf subproblems} whose solution is stored in the dynamic programming table, (ii) the {\bf optimal substructure property} of the problem, and (iii) the {\bf overlap} among subproblems. Then, if needed, we transform the recurrence relation of the dynamic programming algorithm in such a way that it matches the pattern of \cref{eq:dynamic-recurrence-generic}, and we define the dependency set $S_{i_1,i_2,\dots,i_k}$, the generating set $C_{i_1,i_2,\dots,i_k}$, the dependency index $h$, and the function $f_{\mathcal P}$. 
Also, we bound the average degree of the dependency digraph $G_{\mathcal{P}}(I)$ in terms of the size of the instances.
Further, we show that the dynamic programming algorithm is simple.
Finally, we bound the time complexity of the functions $f_{\mathcal P}$, $f_C$, and $\gamma$. Altogether, this allows us to the establish our quantum speedups via \cref{thm:quantum-dp-lemma}.

\begin{figure}[tb!]
\begin{subfigure}[t]{0.475\textwidth}
        \centering
\includegraphics[page=2,height=0.7\textwidth]{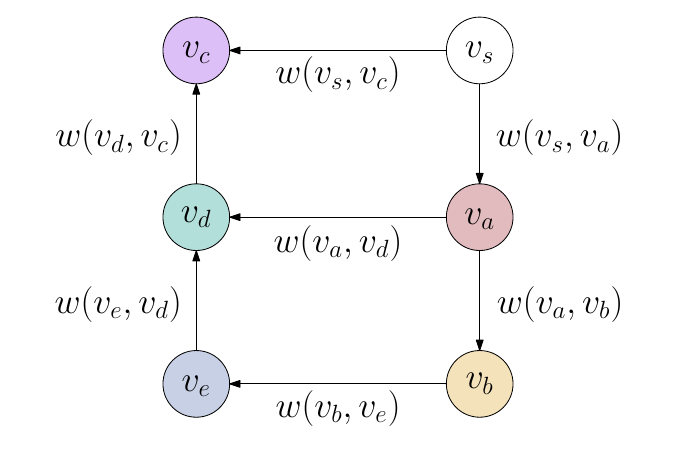} 
\centering 
        \caption{An instance $I=(G,w,s=v_s)$ of SSSP.}
        \label{fig:1a}
    \end{subfigure}
\begin{subfigure}[t]{0.475\textwidth}
\includegraphics[page=18,height=0.7\textwidth]{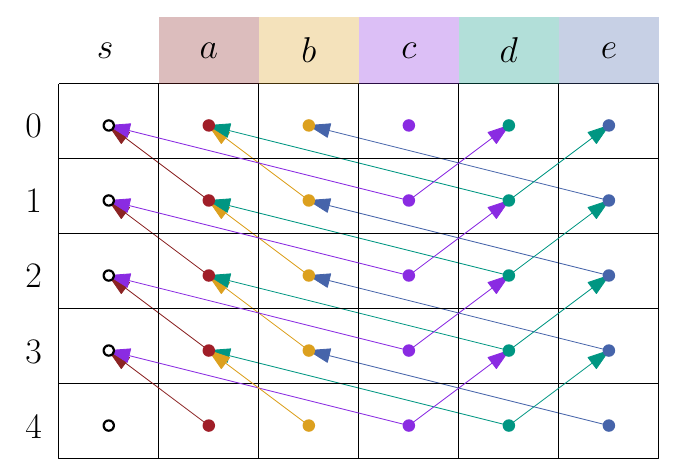} 
\centering 
        \caption{The dependency graph $\depgr$ of~$I$ and the dynamic programming table $D$ for instance $I$. Each node of $\depgr$ is placed inside the corresponding entry of $D$.}
        \label{fig:1b}
    \end{subfigure}

\caption{\centering Illustrations for problem SSSP.}
\label{fig:bellman-dep-graph}
\end{figure}

Let $G=(V,E)$ be a weighted $n$-vertex $m$-edge digraph, where each arc $e = uv \in E$, directed from $u$ to $v$, has an associated weight $w(e)$. Let $p = (u_0,u_1,\dots,u_k)$ be a directed path from $u_0$ to $u_k$. 
The \emph{weight} of $p$ is 
$\sum_{(u_iu_{i+1}) \in E(p)} w(u_i u_{i+1})$, and the \emph{length} of $p$ is~$|E(p)| = k$. 
A \emph{shortest path} from $u$ to $v$ is a minimum-weight directed path from $u$ to $v$. 
The \emph{shortest path distance} from $u$ to $v$ is the weight of a shortest path from $u$ to $v$.
 
Next, we define the {\sc Single-Source Shortest Paths} problem (see \cref{fig:1a} for the illustration of an instance of this problem).

\problemQuestionOutput{\sc Single-Source Shortest Paths (SSSP)}%
{A weighted digraph $G=(V,E)$, a function $w: E \rightarrow \mathbb R$, and a source vertex $s \in V$.}%
{The shortest-path distances between $s$ and $v$, for all $v \in V$.}

\medskip\noindent
\textsc{Subproblems}: 
For all $i \in [n]$ and for all $v_j \in V$,
a subproblem is defined as finding the smallest weight of a directed path, of length at most $i$, from the source vertex $s$ to the vertex~$v_j$. Since the maximum length of a shortest path is at most $n-1$, we have $n^2$ subproblems.

\medskip\noindent
\textsc{Optimal substructure}: 
Let $D$ be a dynamic programming table of size $n \times n$, whose entries $D[i][j]$ store the 
 smallest weight of a directed path, of length at most $i$, 
from $s$ to $v_j$, for all $i \in [n]$ and for all $v_j \in V$.
If the smallest-weight path $p_{s,v_j}$, of length at most $i$, from $s$ to $v_j$ 
uses $uv_j$ as its last arc, then the subpath of $p_{s,v_j}$ from $s$ to $u$ must be a smallest-weight path, of length at most $i-1$, from $s$ to $u$. 
Consider a smallest-weight path $P$, of length at most $i$, from $s$ to $v_j$, whose weight is stored in $D[i][j]$:
\begin{itemize}
    \item If $P$ is of length at most $i-1$, then $D[i][j] = D[i-1][j]$;
    \item If $P$ is of length (at most) $i$ and its last edge is $v_k v_j$, then $D[i][j]= D[i-1][k] + w(v_k v_j)$.
\end{itemize}

\medskip\noindent
\textsc{Overlapping subproblems}: 
The value $D[i-1][k]$ must be accessed to compute all entries $D[i][y]$, where $v_k v_y \in E$.

\smallskip
\noindent
The above yields the following dynamic programming algorithm for SSSP, due to Bellman and Ford~\cite{e7a0cc48-246d-3ba8-9112-de6faa7aa794, Dantzig1997}. The entries of $D$ can be computed as follows. In the base case, $i=0$, we have that
$D[i][j] = \infty$ for all $ v_j \neq s$ and that
$D[i][j] = 0$ if  $v_j = s$.
In the  recursive case, $i>0$, we have that 
$D[i][j] = \min(D[i-1][j],\min_{v_kv_j\in E}(D[i-1][k]+w(v_kv_j)))$.
Note that, this recursive equation can be rewritten as:

$$
D[i][j] = \min_{v_kv_j\in E}(\min(D[i-1][j], D[i-1][k]+w(v_kv_j))),$$
which matches  the pattern of \cref{eq:dynamic-recurrence-generic}. In particular, we 
have that: $S_{i,j}=\{(i-1,k): v_k v_j\in E\}$, $C_{i,j} = \{\{(i-1,k)\}: v_kv_j\in E\}$, $h=1$, and $f_{\mathcal P}(i,j,\{(i-1,k)\}) = \min(D[i-1][j],D[i-1][k]+w(v_k v_j))$. Note that, SSSP is a simple problem, since each entry of $C_{i,j}$ stems from a distinct entry of $S_{i,j}$.
Also observe that, the values in $D$ are bounded by $W = (n-1) \max_{v_i v_j \in E} w(v_i v_j)$. As in \cite{DBLP:journals/jcss/Gabow85, DBLP:journals/siamcomp/GabowT89,DBLP:journals/siamcomp/Goldberg95}, we assume that the weights defined by $w$ are in $O(n^c)$, where $c$ is a constant. This implies that %
each of the entries of $D$ can be represented using
$\log W \in O(\log n )$ bits.
The time $T_{\mathcal P}$, $T_{f_C}$, and $T_\gamma$ needed for computing quantumly $f_{\mathcal P}$, $f_C$, and 
$\gamma$
is then $O(\log W) = O(\log n)$, 
$O(\log n)$,
and 
$O(\log n)$, respectively. Thus, we have that the time
$T' = T_{f_C} + T_\gamma$ (see the statement of \cref{th:minmax}) is~in~$O(\log n)$.

Finally, we bound the average degree of $\depgr$. Recall that $I = (G, w,s)$. Note that, $|V(\depgr)|$ is the number of subproblems, that is, $n^2$.
The $n$ nodes of $V(\depgr)$ corresponding to the subproblems $D[0][\cdot]$ have outdegree $0$.
Instead, by the definition of $C_{i,j}$, each node $n_{i,j}$ of $\depgr$ corresponding to a subproblem $D[i][j]$ has outdegree equal to $\deg_{out}(v_j)$; refer to \cref{fig:1b}. Therefore, for any $i \in \{1,\dots,n-1\}$,  we have that $\sum_{v_j \in V} \deg_{out}(n_{i,j}) = m$.
It follows that the average degree $\delta$ of $\depgr$ is equal to $\frac{2m(n-1)}{n^2} = \frac{2m}{n} - \frac{2m}{n^2} < \frac{2m}{n}$.

Finally, we compare the current-best classical time bound for the SSSP problem with our quantum bound. We remark that the running time of the Bellman-Ford algorithm is $O(n^3)$ time. The current-fastest classical algorithm for SSSP, due to Huang, Jin, and Quanrud~\cite{DBLP:conf/soda/HuangJQ25}, runs in $\tildeO(mn^{4/5})$ time. 
Our quantum dynamic programming version of the Bellman-Ford algorithm instead runs in $\tildeO(n\sqrt{nm})$ time, improving upon~\cite{DBLP:conf/soda/HuangJQ25} when $m \in \Omega(n^{\frac{7}{5}}) = \Omega(n^{1.4})$. 

We remark that, as in the classical setting, by explicitly computing the dynamic programming table, our algorithm allows to detect a negative cycle in $\bigO(m)$ time.

\section{Further Applications of the Quantum Dynamic Programming Framework}\label{se:appendix-uniform}

In this section, we describe several problems to which our quantum dynamic programming framework can be applied; refer to \cref{tab:complexities}.
For each of these problems ${\mathcal P}$, we demonstrate the applicability of \cref{thm:quantum-dp-lemma} as follows. 
First, we present the corresponding classical dynamic programming algorithm. In particular, we describe (i) the {\bf subproblems} whose solution is stored in the dynamic programming table, (ii) the {\bf optimal substructure property} of the problem, and (iii) the {\bf overlap} among subproblems. Then, if needed, we transform the recurrence relation of the dynamic programming algorithm in such a way that it matches the pattern of \cref{eq:dynamic-recurrence-generic}, and we define the dependency set $S_{i_1,i_2,\dots,i_k}$, the generating set $C_{i_1,i_2,\dots,i_k}$, the dependency index $h$, and the function $f_{\mathcal P}$. 
Also, we bound the average degree of the dependency digraph $G_{\mathcal{P}}(I)$ in terms of the size of the instances.
Further, we show that the dynamic programming algorithm is simple.
Finally, we bound the time complexity of the functions $f_{\mathcal P}$, $f_C$, and $\gamma$. Altogether, this allows us to the establish our quantum speedups via \cref{thm:quantum-dp-lemma}.

\subsection{Computational Geometry}

In the following, we will describe several problems within the field of computational geometry.

\subsubsection{Minimum-Weight Triangulation of Convex Polygon}\label{ssc:MWT}

A \emph{triangulation} of a set $S$ of $n$ points in the plane is a maximal plane $n$-vertex graph whose vertices are mapped to the points in $S$.

\problemQuestionOutput{\sc Minimum-weight triangulation of convex polygon (MWT)}%
{A convex polygon $P$ defined by its vertices listed in clockwise order $(v_1,v_2,\dots,v_n)$.}%
{The weight of a triangulation $T$ of $S = \{v_1,v_2,\dots,v_n\}$ of minimum \emph{weight}, defined as the sum of the Euclidean distances of the endvertices of all the edges of $T$.}

\medskip\noindent
Let $T$ be a triangulation of a set of points $S = \{ v_1,\dots,v_n\}$. We denote by $d(v_i,v_j)$ the Euclidean distance between vertices $v_i$ and $v_j$ of $T$.

\medskip\noindent
\textsc{Subproblems}: A subproblem is defined as finding the weight of a minimum-weight triangulation for polygon $(v_i,v_{i+1},\dots,v_{j})$ where $v_i,v_{i+1},\dots,v_{j}$ is a subsequence of (the circular sequence) $v_1,v_2,\dots,v_n$. 
This results in $\frac{n(n-1)}{2}$ subproblems.

\medskip\noindent
\textsc{Optimal substructure}: Let $D$ be a dynamic programming table of size $n \times n$, whose entries $D[i][j]$ store the weight of the minimum-weight triangulation for polygon $(v_i,v_{i+1},\dots,v_{j})$. 
In a minimum-weight triangulation $T$, the edge $(v_1,v_n)$ is part of a triangle with a third vertex $v_r$. The total weight of $T$ is given by:
\begin{equation}\label{eq:MWT}
D[1,n] = \min_{1<r<n} \{d(v_1,v_n) + D[1,r]+D[r,n]\}
\end{equation}

\medskip\noindent
\textsc{Overlapping subproblems}: The value $D[i][j]$ must be accessed to compute all entries $D[k][l]$, where $k<i$ and $j<l$.

\smallskip
\noindent Expanding \cref{eq:MWT} yields the following dynamic programming algorithm for MWT.

\thm{(MWT)}{

\begin{description}
\item[\sc Base case:] If $j= (i +1) \mod (n+1)$, then we have:
\begin{equation*}D[i][j]=d(v_i,v_j)\end{equation*}

\item[\sc Recursive case:] We use the notation $i \prec_r j$ to denote the fact that $v_r$ appears between $v_i$ and $v_j$ when clockwise traversing the boundary of $P$ from $v_i$ to $v_j$.
To compute $D[i][j]$, we need to solve $\bigO(|j-i|)$ subproblems of the form $D[i][r]$ and $D[r][j]$, for all $r$ such that $i \prec_r j$. %
In fact, we have:

\begin{equation}\label{eq:DP-MWT}
D[i][j]=\min_{r: i \prec_r j} \{d(v_i,v_j) + D[i][r] + D[r][j]\}    
\end{equation}
\end{description}
}

Clearly, \cref{eq:DP-MWT} matches the pattern of \cref{eq:dynamic-recurrence-generic}. In fact, we can set $S_{i,j}=\{(i,r),(r,j), r : i \prec_r j\}$, $C_{i,j} = \{\{(i,r),(r,j)\}, r :i \prec_r j\}$, $h=2$, and $f_{\mathcal P}(i,j,\{(i,r),(r,j)\}) = d(v_i,v_j) + D[i][r] + D[r][j]$. Note that, MWT is a simple problem, since each entry of $C_{i,j}$ stems from a distinct entry of $S_{i,j}$. 
The time $T_{\mathcal P}$, $T_{f_C}$, and $T_\gamma$ needed for computing quantumly $f_{\mathcal P}$, $f_C$, and 
$\gamma$
is then $O(\log W) = O(\log n)$, 
$O(\log n)$,
and 
$O(\log n)$, respectively. Thus, we have that the time
$T' = T_{f_C} + T_\gamma$ (see the statement of \cref{th:minmax}) is~in~$O(\log n)$.

Finally, we bound the average degree of $\depgr$. Recall that $I = P = (v_1, v_2, \dots, v_n)$. Note that, $|V(\depgr)|$ is the number of subproblems, that is, $n^2$.
Note that, the $n$ nodes of $V(\depgr)$ corresponding to the subproblems $D[0][\cdot]$ have outdegree $0$.
Instead, by the definition of $C_{i,j}$, each node $n_{i,j}$ of $\depgr$ corresponding to a subproblem $D[i][j]$ has outdegree equal to $|j-i|$. Therefore $\sum_{i=1}^{n-1}\sum_{j=i+1}^{n} |j-i| = \frac{n(n-1)(n+1)}{6}$.
It follows that the average degree $\delta$ of $\depgr$ is equal to $\frac{n(n-1)(n+1)2}{6} / \frac{n(n-1)}{2} = \frac{2(n+1)}{3} < n$.

Finally, we compare the current-best classical time bound \cite{KLINCSEK1980121} for the MWT problem with our quantum bound.

\runningtime{(MWT)}{
\textsc{Classical}: $\bigO(n^3)$ time -- $\bigO(n^2)$ space \cite{KLINCSEK1980121}
\\
\\
\textsc{Quantum}: $\tildeO(n^2\sqrt{n})$ time -- $\bigO(n^2)$ space
}

\subsubsection{Segmented Least Squares Problem}\label{ssc: SLS}

Given a set $P$ of $n$ points in the plane, $(x_1, y_1), (x_2, y_2), \dots, (x_n, y_n)$, with $x_1 < x_2 < \dots < x_n$, and a line $l$ defined as $y_l = a_lx_l+b_l$, the sum of the squared error $e_l$ of the line $l$ with respect to $P$ is defined as follows:
\begin{equation*}
    e_l = \sum_{i=1}^{n} (y_i -a_lx_i -b_l)^2
\end{equation*}

\problemQuestionOutput{\sc Segmented Least Squares Problem (SLS)}%
{A set $P$ of $n$ points in the plane, $(x_1, y_1), (x_2, y_2), \dots, (x_n, y_n)$, with $x_1 < x_2 < \dots < x_n$, a constant penalty $C$ for each line}%
{A partition of $P$ into $m$ line segments that minimizes both $\sum_{s=1}^{m} e_m$ and $Cm$}

\medskip\noindent
Let $D[j]$ denote the minimum penalty for the first $j$ points, $p_1, p_2, \dots, p_j$. Let $e_{i,j}$ denote the minimum error of the best-fit line for the segment containing points $p_i, p_{i+1}, \dots, p_j$. Let $C$ be the cost of adding a new segment (a constant penalty for each segment).

\medskip\noindent
\textsc{Subproblems}: A subproblem is defined as finding the optimal segmented least squares solution for the first $j$ points, where $j$ ranges from $1$ to $n$ (the total number of points).
Hence, we have $\bigO(n)$ subproblems.

\medskip\noindent
\textsc{Optimal substructure}: The optimal solution for $D[j]$ can be constructed from the optimal solutions of smaller subproblems. Specifically, to compute $D[j]$, all possible positions $i$, with $1\leq i \leq j$, where the last segment could begin must be considered.
For each possible starting point $i$ of the last segment, then the optimal solution for the first $j$ points consist of:
\begin{itemize}
    \item The error $e_{i,j}$ of the best-fit line for the segment $p_i, p_{i+1}, \dots, p_j$,
    \item The cost $C$ of adding a new segment, and
    \item The optimal solution for the first $i-1$ points, that is $D[i-1]$.
\end{itemize}
The optimal solution is the minimum over all possible choices of $i$.

\medskip\noindent
\textsc{Overlapping subproblems}: The value $D[j]$ must be accessed to compute all entries $D[k]$, where $j\leq k$.

\smallskip

\noindent This yields the following dynamic programming algorithm for SLS.

\thm{(SLS)}{

\begin{description}
\item[\sc Base case:] For $j = 0$ there are no points and no penalty
\begin{equation*}
D[0] = 0
\end{equation*}

\item[\sc Recursive case:] For $j > 0$, compute:

\begin{equation}\label{eq:DP-SLS}
D[j] = \min_{1 \leq i \leq j} (e_{i,j} + C + D[i-1] )
\end{equation}

\end{description}
}

Clearly, \cref{eq:DP-SLS} matches the pattern of \cref{eq:dynamic-recurrence-generic}. In fact, we have that $S_{j}=\{i: 1\leq i\leq j\}$, $C_{j} = \{\{i\}, 1\leq i \leq j\}$, $h=1$, and $f_{\mathcal P}(i,\{j\}) = e_{i,j} + C + D[i-1]$. Note that, SLS is a simple problem, since each entry of $C_{j}$ stems from a distinct entry of $S_{j}$. The running time for computing $f_{\mathcal P}$, $f_C$, and $_\gamma$ is $\bigO(\log n)$, $\bigO(\log n)$, and $\bigO(\log n)$, respectively. Thus, we have that the time $T'=T_{f_C} +T_{\gamma}$ (see statement of \cref{th:minmax}) is in $\bigO(\log n)$.

Finally, we bound the average degree of $\depgr$. Recall that $I = (P,c)$. Note that, $|V(\depgr)|$ is the number of subproblems, that is, $n$.
Note that, the single node of $V(\depgr)$ corresponding to the subproblem $D[0]$ has outdegree $0$.
Instead, by the definition of $C_{j}$, each node $n_{j}$ of $\depgr$ corresponding to a subproblem $D[j]$ has outdegree equal to $j$. Therefore, we have that $\sum_{j=1}^{n} \deg_{out}(n_{j}) = \frac{n(n+1)}{2}$.
It follows that the average degree $\delta$ of $\depgr$ is equal to $\frac{2n(n+1)}{2}/n = n+1$.

Finally, we compare the current-best classical time bound~\cite{DBLP:journals/cacm/Bellman61a} for the SLS problem with our quantum bound.

\runningtime{(SLS)}{
\textsc{Classical}: $\bigO(n^2)$ time -- $\bigO(n)$  space \cite{DBLP:journals/cacm/Bellman61a}
\\
\\
\textsc{Quantum}: $\tildeO(n\sqrt{n})$ time -- $\bigO(n)$ space
}

\subsubsection{One-Sided Multi-Criteria Boundary Labeling}\label{ssc: MCBL}

Boundary labeling~\cite{bksw-blmearm-07,bnn-eltts-19a,DBLP:journals/jgaa/BenkertHKN09} is a problem in computational geometry motivated from applications in map labeling, where the input consists of a set of $n$ points within a rectangle $R$, e.g., a set of point features in a city map. Each point comes with a specific \emph{label}, e.g., the feature names, which are modeled as unit-height bounding boxes of these names. In the one-sided setting, we further have $n$ uniformly-spaced positions for the labels along one side $a$ of $R$ such that no two labels overlap. A \emph{labeling} $L$ of such an instance consists of a matching between the $n$ feature points and the $n$ labels together with a specific reference point on the label boundary touching $a$ such that each matching pair is connected by \emph{leader} of a specific shape, which indicates the association between a feature and its corresponding label, see \cref{fig:polabeling}. If no two leaders cross, we call $L$ \emph{crossing-free}. Here, we consider orthogonal one-bend leaders (called $po$-leaders~\cite{bnn-eltts-19a}), starting with a segment (called the \emph{hand}) \emph{p}arallel to $a$ in the feature point and ending with a segment (called the \emph{arm}) \emph{o}rthogonal to $a$ in the reference point. The quality of a leader $l$ can be measured by a badness function $bad(l)$ assigning a non-negative badness score to $l$; a common example is the length of the leader or whether it has a bend or not. This leads to the following optimization problem:

\begin{figure}
    \centering
    \includegraphics{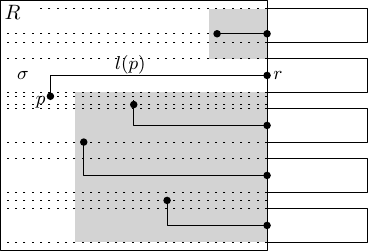}
    \caption{Example of a one-sided boundary labeling instance and how the leader $l(p)$ splits it into two subinstances.}
    \label{fig:polabeling}
\end{figure}

\problemQuestionOutput{\sc One-Sided Multi-Criteria Boundary Labeling (OSMCBL)~\cite{DBLP:journals/jgaa/BenkertHKN09}}%
{A set $P$ of $n$ points in an axis-aligned rectangle $R$ and $n$ candidate positions for disjoint unit-height labels on one side of $R$, and a given badness function $bad()$.}%
{The minimum total badness $\sum_{l\in L} bad(l)$ overall crossing-free labelings $L$.}

\medskip\noindent
For defining subproblems, assume that $a$ is the right side of $R$ as shown in \cref{fig:polabeling} and consider the leftmost point $p$ of $P$. The leader $l(p)$ connecting to a reference point $r$ splits the instance $I$ into two subinstances, one consisting of the points above the arm of $l(p)$ and the labels above $r$ and one consisting of the points below the arm of $l(p)$ and the labels below $r$. This is due to the fact that the labeling must be crossing free and $l(p)$ connects to the leftmost point of $I$. Obviously the number of feature points and labels in each of the two subsinstances must be equal in order to admit a solution.

\medskip\noindent
\textsc{Subproblems}: $R$ is subdivided into $\bigO(n)$ horizontal strips, defined by the horizontal lines passing through the points in $P$ and the horizontal edges of the labels. To determine an optimal crossing-free labeling all possible choices of the strip $\sigma$ for placing the arm of the leader of $p$ must be considered. Consequently, the subproblems are defined by all possible subsets of contiguous strips between two specified strips $i$ and $j$, which are occupied by the arms of leaders of larger subproblems. This approach results in $\bigO(n^2)$ distinct subproblems. We remark that not all of these subproblems admit a valid solution, in which case the total badness of such a subproblem is defined as $\infty$.

\medskip\noindent
\textsc{Optimal substructure}: An optimal crossing-free labeling $L$ is determined by evaluating, for each possible choice of strip $\sigma$, the optimal leader to $p$ in terms of the badness function whose arm lies within $\sigma$, and recursively computing the optimal crossing-free labelings above and below $\sigma$.

\medskip\noindent
\textsc{Overlapping subproblems}: During the computation and decomposition of subproblems by placing leaders for the points from left to right, subproblems being defined by the same subset of strips can occur several times.

\smallskip

\noindent This yields the following dynamic programming algorithm for OSMCBL with $po$-leaders.

\thm{(OSMCBL with $po$-leaders)}{
Let $D[i][j]$ store the value of the minimum total badness for labeling points in the region between two horizontal strips $i$ (bottom) and $j$ (top). The entries of $D$ can be computes as follows:

\begin{description}
\item[\sc Base case:] If there are no further labels between strips $i$ and $j$, we have:  

$$D[i][j] = 0$$

\item[\sc Recursive case:] Let $S(i, j)$ represent the set of strips lying strictly between $i$ and $j$, excluding $i$ and $j$ themselves. Let $p(i, j)$ denote the $k$-th rightmost point within $S(i, j)$, where $k$ is the number of labels entirely covered by $S(i, j)$. 
It is important to note that we only consider {\em feasible choices} of $\sigma$, where the number of labels above (respectively below) $\sigma$ matches the number of points above (respectively below) it. To compute $D[i][j]$, we use the following recurrence: 

$$D[i][j] =
\min_{i < \sigma < j} \left( \text{bad}(l^*(p(i, j), \sigma)) + D[i][\sigma] + D[\sigma][j] \right)$$

where $l^*(p(i, j), \sigma)$ is the optimal leader for the leftmost relevant point $p(i,j)$ in the subproblem, $\sigma$ is a horizontal strip for the placement of the leader’s arm, and bad$(l^*(p(i, j), \sigma))$ is a function that returns the badness score of $l^*(p(i, j), \sigma)$; notice that if $\sigma$ is not a feasible strip we have bad$(l^*(p(i, j), \sigma)) = \infty$ to discard this solution.
\end{description}
}

Before arguing that we can apply \cref{thm:quantum-dp-lemma} we show that the point  $p(i,j)$ can be computed in a quantum way in $\tildeO(\sqrt{n})$ time. %
The set of strips $S(i,j)$ defines an interval of y-coordinates and we can check for each point in $P$ in constant time whether it lies within $S(i,j)$. By putting all $n$ points in superposition we can obtain and mark the points in $S(i,j)$ in constant time. It remains to find the $k$-th rightmost point among the marked ones, which can be done in $\tildeO(\sqrt{n})$ time by iteratively sampling a point from this set and counting the marked points to its right, similarly to median finding using quantum counting and Grover's search~\cite{DBLP:journals/corr/quant-ph-9607014,DBLP:conf/stoc/Grover96}. If there are $k' < k-1$ to the right, we discard them and search for the $(k-k')$-th point in the remaining set. Otherwise we discard all points to the left and repeat the search procedure.

Clearly the recurrence for $D[i][j]$ matches the pattern of \cref{eq:dynamic-recurrence-generic}. In fact, we can set $S_{i,j}=\{(i,\sigma),(\sigma,j): i< \sigma < j\}$, $C_{i,j} = \{\{(i,\sigma),(\sigma,j)\}: i< \sigma < j\}$, $h=2$, and $f_{\mathcal P}(i,j,\{(i,\sigma),(\sigma,j)\}) = \text{bad}(l^*(p(i, j), \sigma)) + D[i][\sigma] + D[\sigma][j]$. 
The definition of $C_{i,j}$ implies that OSMCBL is a simple problem. 
The times $T_{f_C}$ and $T_\gamma$ for computing $f_C$ and $\gamma$ are $O(\log n)$ each. Finally,
assuming that the badness score of any leader is polynomially bounded in the input size (except the special symbol $\infty$ representing infeasible instances), the running time $T_{\mathcal P}$ for computing $f_{\mathcal P}$ is $O(\log n)$. The most critical part is the evaluation of the function bad$(l^*(p(i, j), \sigma))$, since it also includes checking whether $\sigma$ is feasible or not. The score of the leader itself is usually a function that can be evaluated in constant time, as it typically includes only the length of the arm and hand of a leader and whether the leader has a bend or not. The feasibility can be checked in $O(\log n)$ time via two orthogonal range counting queries. For this we use a classical global orthogonal range counting data structure storing the point set $P$, e.g., a range tree which takes $O(n \log n)$ construction time and space and can perform counting queries in $O(\log n)$ time. For the two candidate subproblems $D[i][\sigma]$ and $D[\sigma][j]$ we count the number of points in the rectangle defined by the two strips $i$ and $\sigma$ (or $\sigma$ and $j$) and the x-coordinates of $p(i,j)$ and the right boundary $a$ of $R$. If these numbers do not match the number of labels entirely covered by $S(i,\sigma)$ (or $S(\sigma,j)$) the strip $\sigma$ is not feasible and we assign bad$(l^*(p(i, j), \sigma)) = \infty$. With $\delta = O(n)$ we can now apply \cref{thm:quantum-dp-lemma} and obtain the following speedup.

\runningtime{(OSMCBL with $po$-leaders)}{
\textsc{Classical}: $\bigO(n^3)$ time -- $\bigO(n^2)$ space~\cite{DBLP:journals/jgaa/BenkertHKN09}
\\
\\
\textsc{Quantum}: $\tildeO(n^2\sqrt{n})$ time -- $\bigO(n^2)$ space
}

Benkert et al.~\cite{DBLP:journals/jgaa/BenkertHKN09} also present a dynamic programming algorithm for computing labelings with other leader shapes, namely $do$-leaders with fixed-slope diagonal hand segments. Since the construction of each table entry is similar to the one just described, differing only in the table size, we have described only the $po$-leader case above. However, the same framework can naturally be extended to the $do$-leader case, achieving the same quadratic speedup for the computation of each entry.

\subsection{Combinatorics and Number Theory}

In the following, we will describe problems that fall within the field of combinatorics and number theory.

\subsubsection{RNA Secondary Structure}\label{ssc: RNA-SS}

Let $\{A,U,C,G\}$ be the set of bases forming a single-stranded RNA molecule.
The set of pairs formed by a single-stranded RNA molecule that pairs with itself is called the secondary structure.

Let $S$ be the secondary structure, which must satisfy the following constraints:
    \begin{enumerate}
        \item Pairs are separated by at least four bases.
        \item Pairs consist of either $\{A, U\}$ or $\{C, G\}$.
        \item $S$ is a matching (no base is paired more than once).
        \item No two pairs cross (the structure is non-crossing).
    \end{enumerate}

\problemQuestionOutput{\sc RNA Secondary Structure (RNA-SS)}%
{A single-stranded RNA molecule $B = b_1b_2\dots b_n$, where each $b_i \in \{A, C, G, U\}$}%
{The maximum number of base pairs that can be obtained by a secondary structure $S$ while satisfying the constraints}

\medskip\noindent
\textsc{Subproblems}: For all $1 \leq i < j \leq n$, a subproblem is defined as finding the maximum number of base pairs in a secondary structure for the subsequence $b_i\dots b_j$ of $B$.
We have $\bigO(n^2)$ subproblems, as we need to solve for all possible pairs $(b_i, b_j)$ where $1 \leq i < j \leq n$.

\medskip\noindent
\textsc{Optimal substructure}: Let $D$ be a dynamic programming table of size $n\times n$, whose entries $D[i][j]$ store the maximum number of base pairs in a secondary structure for the subsequence $b_i\dots b_j$ of $B$.
The optimal solution for $D[i][j]$ can be determined by considering four cases:
\begin{enumerate}
\item $D[i][j] = D[i+1][j-1] +1$ if $(b_i,b_j)$ is a base pair;
\item $D[i][j] = D[i+1][j]$ if $b_i$ is unpaired;
    \item $D[i][j] = D[i][j-1]$ if $b_j$ is unpaired;
    \item $D[i][j] = 1 + D[i][t-1] + D[t+1][j-1]$ if $b_j$ pairs with some base $b_t$ (where $i \leq t < j-4$).
\end{enumerate}
The optimal solution is the maximum over these four cases.

\medskip\noindent
\textsc{Overlapping substructure}: The value $D[k][l]$ must be accessed to compute all entries $D[i][j]$, where $i\leq l <j-5$ and $i+1<k<j-5$.

\smallskip

\noindent This yields the following dynamic programming algorithm for RNA-SS.

\thm{(RNA-SS)}{

\begin{description}
\item[\sc Base case:] Since pairs must be separated by at least four bases, for all $i$ and $j$ where $j - i \leq 4$
\begin{equation*}D[i][j] = 0\end{equation*}

\item[\sc Recursive case:] For all $i$ and $j$ where $j - i > 4$, compute:
\begin{equation}\label{eq:DP-RNA-SS}
D[i][j] = \max \begin{cases}
     D[i+1][j-1] +1 \\
     D[i+1][j] \\
     D[i][j-1] \\
     \displaystyle\max_{i\leq t\leq j-4} \{1 + D[i][t-1] + D[t+1][j-1]\},\text{ where $(b_j,b_t)$ is a valid pair} \\
     
\end{cases}
\end{equation}
\end{description}
}

Note that, \cref{eq:DP-RNA-SS} can be rewritten as:
$$
D[i][j] = \displaystyle\max_{i\leq t\leq j-4} (\max(D[i+1][j-1] +1, D[i+1][j],
     D[i][j-1],1 + D[i][t-1] + D[t+1][j-1])),
$$
where $(b_j,b_t)$ is a valid pair. Clearly, this equation matches the pattern of \cref{eq:dynamic-recurrence-generic}. In fact, we have that $S_{i,j}=\{(i,t-1),(t+1,j-1): i\leq t\leq j-4\}$, $C_{i,j} = \{\{(i,t-1),(t+1,j-1)\}: i\leq t\leq j-4\}$, $h=2$, and $f_{\mathcal P}(i,j,\{(i,t-1),(t+1,j-1)\}) = 1 + D[i][t-1] + D[t+1][j-1]$. Note that, RNA-SS is a simple problem, since each entry of $C_{i,j}$ stems from a distinct entry of $S_{i,j}$. The running time for computing $f_{\mathcal P}$, $f_C$, and $_\gamma$ is $\bigO(\log n)$, $\bigO(\log n)$, and $\bigO(\log n)$, respectively. Thus, we have that the time $T'=T_{f_C} +T_{\gamma}$ (see statement of \cref{th:minmax}) is in $\bigO(\log n)$.

Finally, we bound the average degree of $\depgr$. Recall that $I = (B)$. Note that, $|V(\depgr)|$ is the number of subproblems, that is, $\bigO(n^2)$.
Note that, the $\bigO(n)$ nodes of $V(\depgr)$ corresponding to the subproblem $D[i][j]$, where $j-i\leq 4$ have outdegree $0$.
Instead, by the definition of $C_{i,j}$, each node $n_{i,j}$ of $\depgr$ corresponding to a subproblem $D[i][j]$ has outdegree equal to $j-i-4$. Therefore, we have that $\sum_{i=1}^{n}\sum_{j=i}^{n-4} (j-i-4) = \frac{n(n^2 - 24n + 131)}{6}$.
It follows that the average degree $\delta$ of $\depgr$ is equal to $\frac{n(n^2 - 24n + 131)}{3}/{n^2}  < n$.

Finally, we compare the current-best classical time bound~\cite{DBLP:books/daglib/0015106} for the RNA-SS problem with our quantum bound.

\runningtime{(RNA-SS)}{
\textsc{Classical}: $\bigO(n^3)$ time -- $\bigO(n^2)$  space \cite{DBLP:books/daglib/0015106}
\\
\\
\textsc{Quantum}: $\tildeO(n^2\sqrt{n})$ time -- $\bigO(n^2)$ space
}

\subsubsection{Rod Cutting}\label{ssc: RC}
\problemQuestionOutput{\sc Rod Cutting (RC)}%
{A rod $r$ of length $n$ and an array $price[\,]$ where $price[i]$ denotes the value of a piece of length $i$}%
{Maximum value obtainable by cutting the rod and selling the pieces}

\medskip\noindent
\textsc{Subproblems}: A subproblem is defined as computing the maximum value obtainable for a rod of length $j$. Since the rod has length $n$, we have $O(n)$ subproblems.

\medskip\noindent
\textsc{Optimal substructure}: Let $D$ be a dynamic programming table of size $n$ whose entries $D[j]$ store the maximum value obtained for a rod of length $j$.
This can be determined by considering all possible ways to make the first cut.

\medskip\noindent
\textsc{Overlapping substructure}: The value $D[i]$ must be accessed for computing all entries $D[j]$, where $j<i$.

\smallskip

\noindent This yields the following dynamic programming algorithm for RC.

\thm{(RC)}{

\begin{description}
\item[\sc Base case:] For a rod of length $0$, the maximum value is $0$:
\begin{equation*}
D[0] = 0
\end{equation*}

\item[\sc Recursive case:] For a rod of length $j > 0$, the maximum value $D[j]$ is given by:
\begin{equation}\label{eq:DP-RC}
D[j] = \max_{1 \leq i \leq j} (p(i) + D[j-i] )
\end{equation}
\end{description}
}

Clearly, \cref{eq:DP-RC} matches the pattern of \cref{eq:dynamic-recurrence-generic}. In fact, we have that $S_{j}=\{i: 1\leq j\}$, $C_{j} = \{\{i\}, 1\leq i \leq j\}$, $h=1$, and $f_{\mathcal P}(i,\{j\}) = p(i) + D[j-i]$. Note that, RC is a simple problem, since each entry of $C_{j}$ stems from a distinct entry of $S_{j}$. The running time for computing $f_{\mathcal P}$, $f_C$, and $_\gamma$ is $\bigO(\log n)$, $\bigO(\log n)$, and $\bigO(\log n)$, respectively. Thus, we have that the time $T'=T_{f_C} +T_{\gamma}$ (see statement of \cref{th:minmax}) is in $\bigO(\log n)$. Finally, we bound the average degree of $\depgr$. Recall that $I = (n, price[\ ])$. Note that, $|V(\depgr)|$ is the number of subproblems, that is, $\bigO(n)$.
Note that, the single node of $V(\depgr)$ corresponding to the subproblem $D[0]$ has outdegree $0$.
Instead, by the definition of $C_{i}$, each node $n_{i}$ of $\depgr$ corresponding to a subproblem $D[i]$ has outdegree equal to $i$, where $1\leq i \leq n$. Therefore, we have that $\sum_{i=1}^{n} \deg_{out}(n_{i}) = \frac{n(n+1)}{2}$.
It follows that the average degree $\delta$ of $\depgr$ is equal to $\frac{n(n+1)}{n} \leq n+1$.

Finally, we compare the current-best classical time bound \cite{DBLP:books/daglib/0023376} for the RC problem with our quantum bound.

\runningtime{(RC)}{
\textsc{Classical}: $\bigO(n^2)$ time -- $\bigO(n)$  space \cite{DBLP:books/daglib/0023376}
\\
\\
\textsc{Quantum}: $\tildeO(n\sqrt{n})$ time -- $\bigO(n)$ space
}

\subsubsection{Largest Divisible Subset in Array}\label{ssc: LDS}
Given $n$ integers, a subset $S\subseteq n$ is called {\em divisible} if for every pair $(x,y)\in S$, either $x$ divides $y$ or $y$ divides $x$.

\problemQuestionOutput{\sc Largest Divisible Subset in Array (LDS)}%
{A sorted array $A$ of integers of $n$ elements}%
{Size of the largest divisible subset $S$}

\medskip\noindent
\textsc{Subproblems}: A subproblem is defined as finding the largest divisible subset $S_i$ ending at $A[i]$. Since there are $n$ integers in $A$, we have $\bigO(n)$ subproblems.

\medskip\noindent
\textsc{Optimal substructure}: Let $D$ be a dynamic programming table of size $n$, whose entries $D[i]$ store the size of the largest divisible subset ending at $A[i]$. Consider the size of the largest divisible set ending at $A[j]$ then $D[j]$ is either itself (if no previous element divides it) or $D[j]+1$.

\medskip\noindent
\textsc{Overlapping substructure}: The value $D[i]$ must be accessed to compute all entries $D[j]$, where $1\leq i \leq j$.

\smallskip

\noindent This yields the following dynamic programming algorithm for RC.

\thm{(LDS)}{

\begin{description}
\item[\sc Base case:] Each element is trivially a divisible subset of size $1$, hence for each element $j \in A$
\begin{equation*}
D[i] = \{A[i]\}
\end{equation*}

\item[\sc Recursive case:] For each $j$, with $1\leq j\leq n-1$:
\begin{equation}\label{eq:DP-LDS}    
D[j] =\max_{i \text{ s.t. } A[i]|A[j]}(1,D[i]+1)
\end{equation}

\end{description}
}

Clearly, \cref{eq:DP-LDS} matches the pattern of \cref{eq:dynamic-recurrence-generic}. In fact, we have that $S_{i}=\{j: A[j] | A[i]\}$, $C_{i} = \{\{j\}: A[j] | A[i]\}$, $h=1$, and $f_{\mathcal P}(i,\{j\}) = (1,D[j]+1)$. Note that, LDS is a simple problem, since each entry of $C_{i}$ stems from a distinct entry of $S_{i}$. The running time for computing $f_{\mathcal P}$, $f_C$, and $_\gamma$ is $\bigO(\log n)$, $\bigO(\log n)$, and $\bigO(\log n)$, respectively. Thus, we have that the time $T'=T_{f_C} +T_{\gamma}$ (see statement of \cref{th:minmax}) is in $\bigO(\log n)$. 

Finally, we bound the average degree of $\depgr$. Recall that $I = (n,A)$. Note that, $|V(\depgr)|$ is the number of subproblems, that is, $n$.
Note that, the $1$ node of $V(\depgr)$ corresponding to the subproblem $D[0]$ have outdegree $0$.
Instead, by the definition of $C_{j}$, each node $n_{j}$ of $\depgr$ corresponding to a subproblem $D[j]$ has outdegree equal to $i$. Therefore, we have that $\sum_{i=1}^{n} \deg_{out}(n_{i}) = \frac{n(n+1)}{2}$.
It follows that the average degree $\delta$ of $\depgr$ is equal to $\frac{n(n+1)}{n} \leq n+1$.

Finally, we compare the current-best classical time bound~\cite{LDS} for the LDS problem with our quantum bound.

\runningtime{(LDS)}{
\textsc{Classical}: $\bigO(n^2)$ time -- $\bigO(n)$ space \cite{LDS}
\\
\\
\textsc{Quantum}: $\tildeO(n\sqrt{n})$ time -- $\bigO(n)$ space
}

\subsubsection{Text Segmentation Problem}\label{ssc: TSP}

Let $\texttt{IsWord()}$ be an algorithm that, given a dictionary $L$ and string $s$ output $\texttt{True}$ if $s\in L$ and $\texttt{False}$ otherwise.

\problemQuestionOutput{\sc Text Segmentation Problem (TextSeg)}%
{A string $A$, a dictionary $L$, and a subroutine $\texttt{IsWord()}$.}%
{A boolean value equals to $\texttt{True}$ if $A$ can be partitioned into a sequence of words $w_1,\dots,w_l\ |\ w_i\in L, \forall i = 1,\dots, l$, $\texttt{False}$ otherwise.}

\medskip\noindent

\medskip\noindent
\textsc{Subproblems}: A subproblem is defined as determined if the suffix $A[i\dots n]$ of $A$ can be partitioned into a sequence of words from $L$. Since $A$ have at most $n$ suffices, we have $n$ subproblems.

\medskip\noindent
\textsc{Optimal substructure}: Let $D$ be a dynamic programming table of size $n$, whose entries $D[i]$ store a boolean value equals to $\texttt{True}$ if $A[i\dots n]$ can be partitioned into a sequence of words contained in the dictionary, $\texttt{False}$ otherwise.
Consider the entry $D[i]$, containing the value relative to the suffix $A[i\dots n]$. It's value is $\texttt{True}$ if there exists an $j>j$ such that $A[i\dots j] \in L$ and $D[j+1] = \texttt{True}$.

\medskip\noindent
\textsc{Overlapping substructure}: The value $D[j]$ must be accessed to compute all entries $D[i]$, where $i<j$.

\smallskip

\noindent This yields the following dynamic programming algorithm for \textsc{TextSeg}.

\thm{(\textsc{TextSeg})}{

\begin{description}

\item[\sc Base case:] If $i > n$, then $D[i] = \texttt{True}$ because an empty string can be trivially partitioned.

\item[\sc Recursive case:] For each position $i$ in $A$, we check all possible ways to split the suffix $A[i\dots n]$ into words:
\begin{equation}\label{eq:DP-TSP}
D[i] = \bigvee_{j=i}^{n} ( \texttt{IsWord}(A[i\dots j]) \wedge D[j+1] )
\end{equation}

\end{description}
}

Clearly, \cref{eq:DP-TSP} matches the pattern of \cref{eq:dynamic-recurrence-generic}. In fact, we have that $S_{i}=\{j: i<j\}$, $C_{i} = \{\{j\}: i<j\}$, $h=1$, and $f_{\mathcal P}(i,\{j\}) = ( \texttt{IsWord}(A[i\dots j]) \wedge D[j+1])$. Note that, TSP is a simple problem, since each entry of $C_{i}$ stems from a distinct entry of $S_{i}$. The running time for computing $f_{\mathcal P}$, $f_C$, and $_\gamma$ is $\bigO(\log n)$, $\bigO(\log n)$, and $\bigO(\log n)$, respectively. Thus, we have that the time $T'=T_{f_C} +T_{\gamma}$ (see statement of \cref{th:minmax}) is in $\bigO(\log n)$.

Finally, we bound the average degree of $\depgr$. Recall that $I = (A,L,\texttt{IsWord()})$. Note that, $|V(\depgr)|$ is the number of subproblems, that is, $n$.
Note that, the $1$ node of $V(\depgr)$ corresponding to the subproblem $D[n]$ have outdegree $0$.
Instead, by the definition of $C_{i}$, each node $n_{i}$ of $\depgr$ corresponding to a subproblem $D[i]$ has outdegree equal to $n-i$. Therefore, we have that $\sum_{i=1}^{n} \deg_{out}(n_{i}) = \frac{n(n-1)}{2}$.
It follows that the average degree $\delta$ of $\depgr$ is equal to $\frac{n(n-1)}{n} \leq n+1$.

Finally, we compare the current-best classical time bound~\cite{DBLP:books/x/E2019} for the \textsc{TextSeg} problem with our quantum bound.

\runningtime{(\textsc{TextSeg})}{
\textsc{Classical}: $\bigO(n^2)$  time -- $\bigO(n)$ space \cite{DBLP:books/x/E2019}
\\
\\
\textsc{Quantum}:  $\tildeO(n\sqrt{n})$ time -- $\bigO(n)$ space
}

\subsubsection{Unbounded Knapsack Problem}\label{ssc: UKP}

\problemQuestionOutput{\sc Unbounded Knapsack (UKP)}%
{A set $N$ of $n$ items, each with a weight $w_i, i=1,\dots,n$ and a value $v_i, i=1,\dots,n$ and a knapsack of capacity $W$}%
{The maximum value that can be achieved by choosing a subset $S\subseteq N$ such that $\sum_{j\in S} p_j \leq W$, where the value is equal to $\sum_{j\in S} v_j$.}

\medskip\noindent

\medskip\noindent
\textsc{Subproblems}: A subproblem is defined as finding the maximum value achievable with a knapsack of capacity $i$, with $i\leq W$. It is evident that we have $W$ subproblems.

\medskip\noindent
\textsc{Optimal substructure}: Let $D$ be a dynamic programming table of size $W$, whose entries $D[i]$ store the maximum value achievable with a knapsack of capacity $i$. Consider a subproblem of capacity $i$ whose value is stored in $D[i]$ and consider each item $j$ such that $w_j< i$. Entry $D[i]$ can be computed by computing the best item value $v_j$ to add, with $w_j<i$, to maximize $D[i-p_j]+v_j$.

\medskip\noindent
\textsc{Overlapping substructure}: The value $D[i]$ must be accessed to compute all entries $D[k]$, where $i<k$ and there exist at least an item $j$ such that $k-w_j = i$.

\smallskip

\noindent This yields the following dynamic programming algorithm for UKP.

\thm{(UKP)}{
\begin{description}
\item[\sc Base case:] If $i = 0$: 
\begin{equation*}
    D[i] = 0
\end{equation*}

\item[\sc Recursive case:] For each capacity $i \leq W$:
\begin{equation}\label{eq:DP-UKP}
D[i] = \max_{j: w_j \leq i} \{ D[i - w_j] + v_j \}.
\end{equation}
\end{description}
}

Clearly, \cref{eq:DP-UKP} matches the pattern of \cref{eq:dynamic-recurrence-generic}. In fact, we have that $S_{i}=\{j: w_j\leq i\}$, $C_{i} = \{\{j\}: w_j\leq i\}$, $h=1$, and $f_{\mathcal P}(i,\{j\}) = D[i - w_j] + v_j$. Note that, UKP is a simple problem, since each entry of $C_{i}$ stems from a distinct entry of $S_{i}$. The running time for computing $f_{\mathcal P}$, $f_C$, and $_\gamma$ is $\bigO(\log n)$, $\bigO(\log n)$, and $\bigO(\log n)$, respectively. Thus, we have that the time $T'=T_{f_C} +T_{\gamma}$ (see statement of \cref{th:minmax}) is in $\bigO(\log n)$.

Finally, we bound the average degree of $\depgr$. Recall that $I = (N,W)$. Note that, $|V(\depgr)|$ is the number of subproblems, that is, $W$.
Note that, the $1$ node of $V(\depgr)$ corresponding to the subproblem $D[0]$ have outdegree $0$.
Instead, by the definition of $C_{i}$, each node $n_{i}$ of $\depgr$ corresponding to a subproblem $D[i]$ has outdegree (at most) equal to $n$. Therefore, we have that $\sum_{i=0}^{W} \deg_{out}(n_{i}) = nW$.
It follows that the average degree $\delta$ of $\depgr$ is equal to $\frac{2nW}{W} \leq 2n$.

Finally, we compare the current-best classical time bound~\cite{DBLP:books/daglib/0017733} for the UKP problem with our quantum bound.

\runningtime{(UKP)}{
\textsc{Classical}: $\bigO(Wn)$ time -- $\bigO(W)$ space \cite{DBLP:books/daglib/0017733}
\\
\\
\textsc{Quantum}: $\tildeO(W\sqrt{n})$ time -- $\bigO(W)$ space
}

\subsection{Optimal Paths}

In the following, we will describe two Optimal-Path problems.

\subsubsection{All-Pairs Shortest Paths via Matrix Multiplication}\label{ssc: APSP_Matrix}

Let $G=(V,E)$ be a weighted $n$-vertex $m$-edge digraph, where each arc $e = uv \in E$, directed from $u$ to $v$, has an associated weight $w(e)$. Let $p = (u_0,u_1,\dots,u_k)$ be a directed path from $u_0$ to $u_k$. 
The \emph{weight} of $p$ is 
$\sum_{(u_iu_{i+1}) \in E(p)} w(u_i u_{i+1})$, and the \emph{length} of $p$ is~$|E(p)| = k$. 
A \emph{shortest path} from $u$ to $v$ is a minimum-weight directed path from $u$ to $v$. 
The \emph{shortest path distance} from $u$ to $v$ is the weight of a shortest path from $u$ to $v$.

\problemQuestionOutput{\sc All-Pairs Shortest Paths (APSP) via Matrix Multiplication}%
{A weighted digraph $G=(V,E)$, a function $w: E \rightarrow \mathbb R$.}%
{The shortest-path distances between every pair of vertices $v_i,v_j\in V\times V$.}

\medskip\noindent

\medskip\noindent
\textsc{Subproblems}: For a fixed $l$, for all $v_i,v_j \in V\times V$, a subproblem is defined as the shortest path distance between vertices $v_i$ and $v_j$ using at most $l$ edges.
We have $\bigO(n^2)$ subproblems.

\medskip\noindent
\textsc{Optimal substructure}: Let $D^l$ be a dynamic programming table of size $n\times n$, whose entries $D^l[i][j]$ store the shortest path distance between vertices $v_i$ and $v_j$ using at most $l$ edges. 
If the shortest path from $v_i$ to $v_j$ use at most $2l$ edges passes through some intermediate vertex $k$, then:
$$
D^{(2l)}[i][j] = \min_{{v_k} \in V} (D^{(l)}[i][k] + D^{(l)}[k][j])
$$
This shows that the shortest path using at most $2l$ edges can be determined from two shorter paths using at most $l$ edges each.

\smallskip\noindent
\textsc{Overlapping substructure}: The value $D^{l}[i][j]$ must be accessed to compute all entries $D^{2l}[g][k]$, where $g\leq i \leq j \leq k$.

\noindent This yields the following dynamic programming algorithm for APSP using matrix multiplication.

\thm{(APSP via Matrix Multiplication)}{

\begin{description}
\item[\sc Base case:] For $l=0$, for each $v_i,v_j \in V\times V$:
\begin{equation*}
D^{l}[i][i] =  \begin{cases}
     0, \text{ if } i=j,\\
     w(i,j) \text{ if } (i,j)\in E,\\
     \infty \text{ otherwise.}
\end{cases}
\end{equation*}

\item[\sc Recursive case:] Compute the shortest paths iteratively using matrix multiplication with a modified operation:

\begin{equation}\label{eq:DP-APSP}
    D^{(2l)}[i][j] = \min_{{v_k} \in V} (D^{(l)}[i][k] + D^{(l)}[k][j])
\end{equation}

\end{description}
}

Clearly, \cref{eq:DP-APSP} matches the pattern of \cref{eq:dynamic-recurrence-generic}. In fact, we have that $S_{i,j}=\{(i,k),(k,j), v_k \in V\}$, $C_{i,j} = \{\{(i,k),(k,j)\},v_k \in V\}$, $h=2$, and $f_{\mathcal P}(i,j,\{(i,k),(k,j)\}) = D^{(l)}[i][k] + D^{(l)}[k][j]$. Note that, APSP is a simple problem, since each entry of $C_{i,j}$ stems from a distinct entry of $S_{i,j}$. 
The time $T_{\mathcal P}$, $T_{f_C}$, and $T_\gamma$ needed for computing quantumly $f_{\mathcal P}$, $f_C$, and 
$\gamma$
is then $O(\log W) = O(\log n)$, 
$O(\log n)$,
and 
$O(\log n)$, respectively. Thus, we have that the time
$T' = T_{f_C} + T_\gamma$ (see the statement of \cref{th:minmax}) is~in~$O(\log n)$.

Finally, we bound the average degree of $\depgr$. Recall that $I = (G,w)$. Note that, $|V(\depgr)|$ is the number of subproblems, that is, $n^2$.
Note that, the $n$ nodes of $V(\depgr)$ corresponding to the subproblems $D[i][j]$, where $i=j$ or $(i,j)\in E$, have outdegree $0$.
Instead, by the definition of $C_{i,j}$, each node $n_{i,j}$ of $\depgr$ corresponding to a subproblem $D[i][j]$ has outdegree equal to $n$. Therefore $\sum_{i=1}^{n}\sum_{j=1}^{n} n = n^3$.
It follows that the average degree $\delta$ of $\depgr$ is equal to $\frac{2n^3}{n^2} \leq  2n$.

Finally, we compare the current-best classical time bound~\cite{DBLP:journals/jda/HanT16} for the APSP problem with our quantum bound.

\runningtime{(APSP)}{
\textsc{Classical}:  $\bigO(n^3 \log \log n / \log ^2 n)$ time -- $\bigO(n^2)$ space \cite{DBLP:journals/jda/HanT16}
\\
\\
\textsc{Quantum}: $\tildeO(n^2\sqrt{n}\log n)$  time -- $\bigO(n^2)$ space
}

\subsubsection{Viterbi Algorithm}\label{ssc: VA}

A \emph{transition probability} $a_{r,s}$ is the probability of transitioning from state $r$ to state $s$. An \emph{emission probability} $b_{s,o}$ is the probability of observing $o$ at state $s$.

\problemQuestionOutput{\sc Viterbi Algorithm (VA)}%
{A hidden Markov model with a set of hidden states $S$, a sequence of $T$ observations $o_0, o_1, \dots, o_{T-1}$, initial state probabilities $\pi_s$, transition probabilities $a$, and emission probabilities $b$}%
{The most likely sequence of states that could have produced the observations}

\medskip\noindent

\medskip\noindent
\textsc{Subproblems}: For each time step $t \in T$ and hidden state $s_j \in S$, a subproblem is defined as probability of the most probable path ending in state $s_j$ at time $t$. Since we need to compute a probability for each state at each time step, the total number of subproblems is $|S|\times |T|$.

\medskip\noindent
\textsc{Optimal substructure}: Let $D$ be a dynamic programming table of size $|S|\times |T|$ whose entries $D[t][j]$ store the probability of the most probable path ending in state $s_j$ at time $t$.
The most probable path to state $s_j$ at time $t$ depends on the most probable paths to all possible previous states at time $t-1$. This means that:
$$
D[t][j]=\max_{i \in S} (D[t-1][i]\cdot a_{s_i,s_j})\cdot b_{s_j,o_t}
$$

where $a_{s_i,s_j}$ is the transition probability from state $s_i$ to state $s_j$ and $b_{s_j,o_t}$ is the emission probability of observing $o_t$ given state $s_j$. The probability of reaching $s_j$ at time $t$ is computed using optimal probabilities from the previous steps.

\medskip\noindent
\textsc{Overlapping substructure}: The value $D[t-1][i]$ must be accessed multiple times across different transitions $a_{i,j}$.

\smallskip

\noindent This yields the following dynamic programming algorithm for VA.

\thm{(VA)}{

\begin{description}
\item[\sc Base case:] Let $\pi_{j}$ be the initial probability of state $s_j$ and let $b_{s_j,o_1}$ be the probability of emitting the first observation from state $s_j$, then for $t = 1$:
$$
D[1][j] =\pi_j\cdot b_{s_j,o_1}
$$

\item[\sc Recursive case:] For $t > 1$:
\begin{equation}\label{eq:DP-Viterbi}
    D[t][j]=\max_{i \in S} (D[t-1][i]\cdot a_{s_i,s_j})\cdot b_{s_j,o_t}
\end{equation}
\end{description}
}

Clearly, \cref{eq:DP-Viterbi} matches the pattern of \cref{eq:dynamic-recurrence-generic}. In fact, we have that $S_{t,j}=\{(t-1,i), s_i \in S\}$, $C_{t,j} = \{\{(t-1,i)\}, s_i \in S\}$, $h=1$, and $f_{\mathcal P}(t,j,\{(t-1,i)\}) = (D[t-1][i]\cdot a_{s_i,s_j})\cdot b_{s_j,o_t}$. Note that, VA is a simple problem, since each entry of $C_{t,j}$ stems from a distinct entry of $S_{t,j}$. 
The time $T_{\mathcal P}$, $T_{f_C}$, and $T_\gamma$ needed for computing quantumly $f_{\mathcal P}$, $f_C$, and 
$\gamma$
is then $O(\log W) = O(\log n)$, 
$O(\log n)$,
and 
$O(\log n)$, respectively. Thus, we have that the time
$T' = T_{f_C} + T_\gamma$ (see the statement of \cref{th:minmax}) is~in~$O(\log n)$.

Finally, we bound the average degree of $\depgr$. Recall that $I = (S,T,\pi_s,a, b)$. Note that, $|V(\depgr)|$ is the number of subproblems, that is, $|S|\times |T|$.
Note that, the $|S|$ nodes of $V(\depgr)$ corresponding to the subproblems $D[1][\cdot]$, have outdegree $0$.
Instead, by the definition of $C_{t,j}$, each node $n_{t,j}$ of $\depgr$ corresponding to a subproblem $D[t][j]$ has outdegree equal to $|S|$. Therefore $\sum_{t=1}^{|T|}\sum_{j=1}^{|S|} |S| = |T||S|^2$.
It follows that the average degree $\delta$ of $\depgr$ is equal to $\frac{2|T||S|^2}{|T||S|} \leq  2|S|$.

Finally, we compare the current-best classical time bound \cite{DBLP:books/daglib/0004298} for the VA problem with our quantum bound.

\runningtime{(VA)}{
\textsc{Classical}: $\bigO(T \times |S|^2)$  time -- $\bigO(T \times |S|)$ space \cite{DBLP:books/daglib/0004298}
\\
\\
\textsc{Quantum}: $\tildeO(T \times |S| \times \sqrt{|S|})$  time -- $\bigO(T \times |S|)$ space
}

\subsection{Quantum Cocke-Younger-Kasami for Membership in Context-Free Language}\label{sc:CYK}

Let $ G $ be a \emph{Context-Free Grammar} (CFG) in \textbf{Chomsky Normal Form (CNF)}. 
A CFG is in CNF if all its production rules are of one of the following forms:
\begin{enumerate*}[label=({\bf \arabic*})]
    \item $ A \rightarrow BC $ where $ A, B, C $ are non-terminals symbols (variables), hence a non-terminal $A$ can be replaced by two other non-terminals symbols,
    \item $ A \rightarrow a $ where $ A $ is a non-terminal and $ a $ is a terminal symbol, hence a non-terminal $A$ can be replaced by a single terminal symbol $a$,
\end{enumerate*}
\noindent
Let $L(G)$ be the language composed by all the strings that can be generated by the grammar $G$.
 
 Next, we define the {\sc Membership in Context-Free Language} problem.

\problemQuestionOutput{\sc Membership in Context-Free Language (MCFL)}%
{A constant size context-free grammar $G$ in CNF and a nonempty string $w=w_1w_2\dots w_n$, where each $w_i$ is a terminal symbol.}%
{A boolean value $\texttt{True}$ if the string $w$ can be derived from the start symbol $S$ of the grammar $G$, i.e. $w \in L(G)$, $\texttt{False}$ otherwise.}

\medskip\noindent
\textsc{Subproblems}: A subproblem is defined as determing whether a substring $w_jw_{j+1}\dots w_l$ can be derived from a non-terminal symbol $A$. Since a string of $n$ symbols has $\binom{n}{2}$ consecutive substrings of length at most $n$, we have $\frac{n(n+1)}{2}$ subproblems.

\medskip\noindent
\textsc{Optimal substructure}: Let $D$ be a dynamic programming table of size $\bigO(n^2)$, whose entries $D[i][j]$ store the set of non-terminals that can generate the substring $w_jw_{j+1}\dots w_{j+i-1}$. Consider the subproblem consisting of the substring $w_jw_{j+1}\dots w_{j+i-1}$, then the set of non-terminals for $D[i][j]$ is obtained as follows:
\begin{itemize}
    \item $D[i][j] = ((B \in D[k][j]) \wedge (C \in D[i-k][j+k]))$, where $1\leq k< i$ and $(A\rightarrow BC) \in G$
\end{itemize}

\medskip\noindent
\textsc{Overlapping subproblems}: The value $D[k][j]$ or $D[i-k][j+k]$ must be accessed to compute all the entries of the form $D[u][l]$, where:
\begin{itemize}
    \item $l=j$ and $k<u$;
    \item $l=j+k$ and $i-k<u$.
\end{itemize}

\smallskip
\noindent
The above yields the following dynamic programming algorithm for MCFL, due to Cocke-Younger-Kasami~\cite{DBLP:journals/iandc/Younger67,kasami1966efficient,cocke1969programming}. 

\thm{(MCFL)}{
The entries of $D$ can be computes as follows:

\begin{description}
\item[\sc Base case:] if $i=1$:
\begin{equation*}
    D[i][j] = 
        \{A\ |\  (A\rightarrow w_j) \in G\}
\end{equation*}

\item[\sc Recursive case:] if $i<j$:
\begin{equation}\label{eq:DP-CYK}
    D[i][j] =\findAll_{1\leq k< i,\  A\rightarrow BC } ( (B \in D[k][j]) \wedge (C \in D[i-k][j+k]) )
\end{equation}
\end{description}
}

Note that this recursive equation matches the pattern of \cref{eq:dynamic-recurrence-generic} of \cref{thm:quantum-dp-lemma}. Observe that, $w$ can be generated from $G$ if and only if $D[n][1]\neq \{\emptyset\}$.
In particular, we 
have that: $S_{i,j}=\{(k,j),(i-k,j+k): 1\leq k< i\}$, $C_{i,j} = \{\{(k,j),(i-k,j+k)\}: 1\leq k< i\}$, $h=2$, and $f_{\mathcal P}(i,j,\{(k,j),(i-k,j+k)\}) = (B \in D[k][j]) \wedge (C \in D[i-k][j+k])$. Note that, MCFL is a simple problem, since each entry of $C_{i,j}$ stems from a distinct entry of $S_{i,j}$. 
The time needed for computing quantumly $f_{\mathcal P}$, $f_C$, and 
$\gamma$
is then $O(1)$, 
$O(\log n)$,
and 
$O(\log n)$, respectively. Therefore, we have that 
$T' \in O(\log n)$. 

\begin{figure}[htp]
\centering
\includegraphics[page=20, width=.30\textwidth]{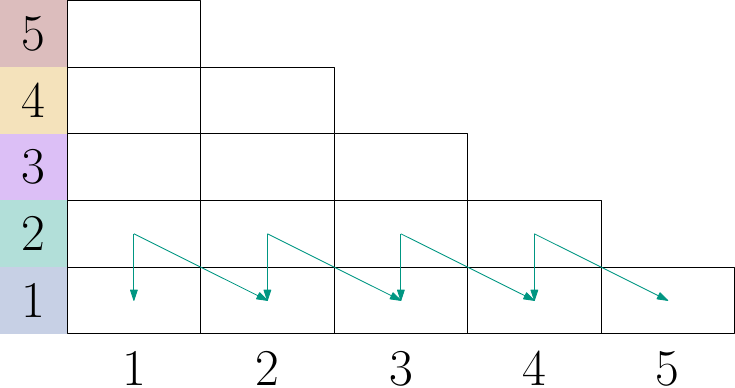}\hfil
\centering
\includegraphics[page=21, width=.30\textwidth]{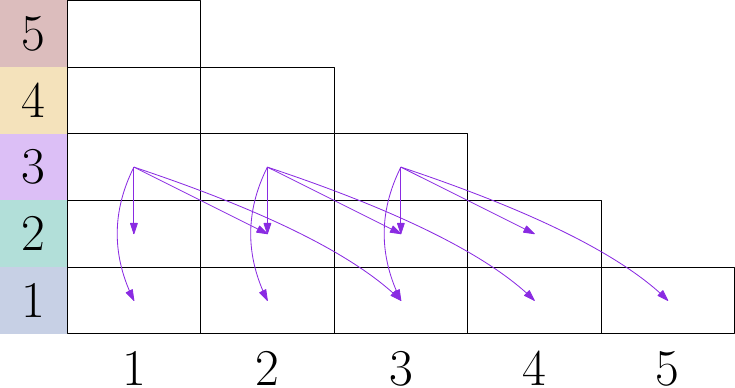}
\\
\centering
\includegraphics[page=22, width=.30\textwidth]{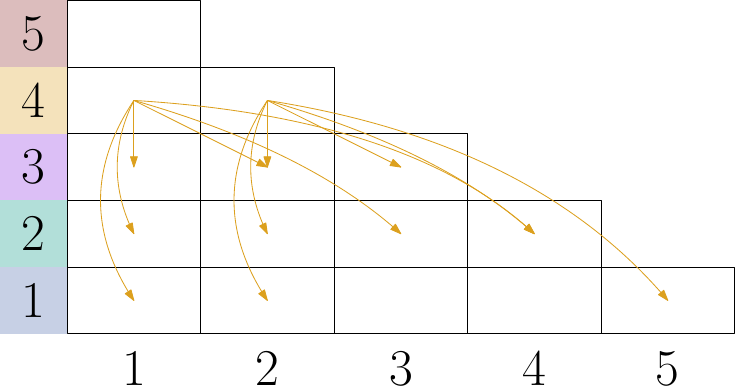}\hfil
\centering
\includegraphics[page=23, width=.30\textwidth]{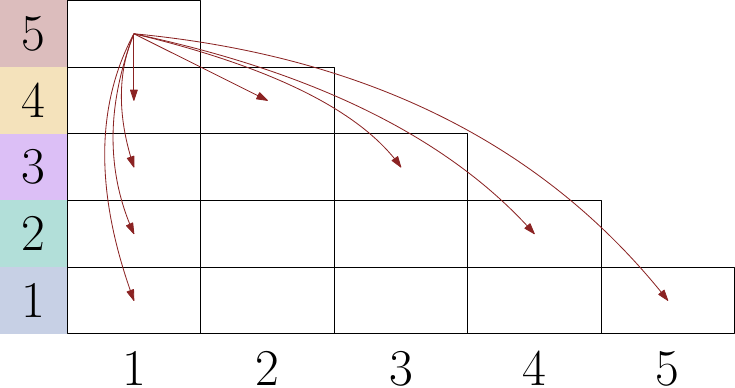}

\caption{\centering Graph $\depgr$ for the dynamic programming algorithm that solves MCFL.}
\label{fig:cyk-table}
\end{figure}

Finally, we bound the average degree of $\depgr$. Recall that $I = (G, w)$. Note that, $|V(\depgr)|$ is the number of subproblems, that is, $\frac{n(n+1)}{2}$.
Note that, the $n$ nodes of $V(\depgr)$ corresponding to the subproblems $D[1][j]$ have outdegree $0$.
Instead, by the definition of $C_{i,j}$, each node $n_{i,j}$ of $\depgr$ corresponding to a subproblem $D[i][j]$ has outdegree equal to $2(i-1)$; refer to \cref{fig:cyk-table}. Therefore, for any $i \in \{1,\dots,n\}$,  we have that $\sum_{j \in [n]} \deg_{out}(n_{i,j}) = 2n(i-1)$. The total number of edges of $\depgr$ is equal to $$\sum_{i=1}^{n} 2n(i-1)=n\sum_{i=1}^{n}(i-1) = 2n(\sum_{i=1}^{n}i-\sum_{i=1}^{n}1)) = 2n(\frac{n(n+1)}{2}-n)) = 2n(\frac{n(n-1)}{2}).$$
It follows that the average degree $\delta$ of $\depgr$ is equal to $$\frac{4n(\frac{n(n-1)}{2})}{\frac{n(n+1)}{2}} = \frac{4n(n-1)}{n+1} = \frac{4n^2}{n+1} - \frac{4n}{n+1} < 4n.$$

Finally, we compare the current-best classical time bound for the MCFL problem with our quantum bound. We remark that the running time of the CYK algorithm is $O(n^3)$ time. 

Valiant demonstrated in 1975 that context-free language recognition can be performed at least as efficiently as Boolean matrix multiplication \cite{DBLP:journals/jcss/Valiant75}. Since Boolean matrix multiplication could be achieved in $\bigO(n^{2.81})$ time using Strassen's algorithm \cite{strassen1970gaussian}, this implied an indirect $\bigO(n^{2.81})$ algorithm for context-free recognition. However, such an algorithm does not allow to directly retrieve the sequence of productions used to derive $w$. The current-fastest classical algorithm for matrix multiplication has a time complexity of $\bigO(n^{2.371552})$ \cite{DBLP:conf/soda/WilliamsXXZ24}, building upon the Coppersmith-Winograd algorithm \cite{DBLP:journals/jsc/CoppersmithW90}. However, the constant factors hidden by the Big O notation are so substantial that these fast matrix multiplication algorithms are impractical for matrix sizes manageable by current computers.

Moreover, our approach significantly simplifies the process. We directly accelerate an ``easy'' algorithm for context-free parsing. In contrast, using matrix multiplication for parsing requires a more complex, three-step process:
\begin{enumerate*}[label=({\bf \arabic*})] \item 
    transforming the parsing instance into a format suitable for matrix multiplication, \item applying the sophisticated (and practically cumbersome) matrix multiplication algorithm described in \cite{DBLP:conf/soda/WilliamsXXZ24}, \item 
    converting the recognition solution (provided by the matrix multiplication) into a parsing solution\footnote{Note that Ruzzo showed parsing is slightly harder than recognition by a logarithmic factor \cite{DBLP:conf/icalp/Ruzzo79}}.
    \end{enumerate*}

Therefore, our algorithm offers a much simpler and more direct approach to context-free parsing compared to methods relying on matrix multiplication.
\runningtime{(CYK)}{
\textsc{Classical}: $\bigO(n^{2.371552})$ \cite{DBLP:conf/soda/WilliamsXXZ24} time -- $\bigO(n^2)$ space 
\\
\\
\textsc{Quantum}: $\tildeO(n^2\sqrt{n})$ time -- $\bigO(n^2)$ space
}

\section{Conclusions and Open Problems}

In this paper, we presented a framework that systematically extends classical dynamic programming algorithms that exhibit specific characteristics in the computation of their dynamic programming table, into accelerated quantum counterparts.
These characteristics pertain to the structure of the {\em dependency set}, specifying the global set of entries involved in computing a table entry, and to the structure of the {\em generating set}, specifying the groups of entries simultaneously involved in providing a candidate value for a table entry.

By leveraging quantum search  primitives, such as {\em find}, {\em findAll}, {\em min}, and {\em max}, and the QRAM, we transform dynamic programming recurrence relations into efficient quantum subroutines. Our approach lowers the computational cost of constructing each entry in the dynamic programming table, often achieving a significant speedup over the best-known classical algorithms. We demonstrate the versatility of this framework by applying it to several well-known problems, including {\sc Single-Source Shortest Paths}. %

This work establishes a foundation for developing more efficient quantum algorithms for a class of dynamic programming problems, paving the way for several research directions.
A natural extension is to investigate the applicability of our framework to dynamic programming algorithms that do not have the above characteristics.
It is also worth investigating whether similar quantum speedups can be achieved for algorithms that iteratively refine estimates of the optimal solution, such as the Floyd-Warshall algorithm.
Finally, a key challenge is optimizing the space complexity of quantum dynamic programming algorithms. Due to the limitations of quantum memory, it is crucial to develop strategies that avoid storing the entire dynamic programming table. This may involve computing entries on-the-fly, leveraging quantum speedups for efficient on-demand computation, and minimizing the number of entries stored in \mbox{superposition}.

\bibliographystyle{abbrv}
\bibliography{bibliography}

\clearpage
\appendix

\end{document}